\documentclass[useAMS]{mn2e}
\usepackage{aasmacros}
\usepackage{times}
\usepackage{graphicx}
\usepackage{amssymb}
\usepackage[usenames,dvipsnames]{color}
\usepackage{url}

\newcommand{\cutt}[1]{}
\newcommand{\newt}[1]{#1}
\newcommand{\movt}[1]{#1}
\newcommand{\reft}[1]{#1}
\newcommand{\npgt}[1]{}
\newcommand{\xtrt}[1]{}

\newcommand{\galform}[0]{{\sc galform}}
\newcommand{\rbulge}[0]{3~kpc}
\newcommand{\gadgetthree}[0]{{\sc gadget-3}}
\newcommand{\gadgettwo}[0]{{\sc gadget-2}}
\newcommand{\subfind}[0]{{\sc subfind}}
\newcommand{\lcdm}[0]{${\Lambda}$CDM}

\newcommand{\fig}[1]{Fig.~\ref{#1}}
\newcommand{\mntab}[1]{Table~\ref{#1}}
\newcommand{\mnsec}[1]{Section~\ref{#1}}
\newcommand{\afe}[0]{[$\rm{\alpha/Fe}$]}
\newcommand{\feh}[0]{[$\rm{Fe/H}$]}

\title[Galactic stellar haloes in the CDM model]{Galactic stellar
  haloes in the CDM model}

\author[A.P. Cooper et al.]{A.P. Cooper$^{1}$\thanks{E-mail:a.p.cooper@durham.ac.uk}, S. Cole$^{1}$, 
  C.S. Frenk$^{1}$, S.D.M. White$^{2}$, J. Helly$^{1}$, A.J. Benson$^{3}$,
  \newauthor G. De Lucia$^{4}$, A. Helmi$^{5}$, A. Jenkins$^{1}$, J.F. Navarro$^{6}$, V. Springel$^{2}$, J. Wang$^{1}$  \\
  $^{1}$Institute for
  Computational Cosmology, Department of Physics, University of
  Durham, South Road, Durham, DH1 3LE, UK \\
  $^{2}$Max-Planck-Institut f\"{u}r Astrophysik, Karl-Schwarzschild-Str. 1, 
  D-85748, Garching, Germany \\
  $^{3}$Mail Code 350-17, California Institute of Technology, Pasadena, CA 91125, U.S.A.\\
  $^{4}$INAF-Osservatorio Astronomico di Trieste, Via G.B. Tiepolo 11, I-34131 Trieste, Italy\\
  $^{5}$Kapteyn Astronomical Institute, University of Groningen, 
  P.O. Box 800, 9700 AV Groningen, Netherlands\\
  $^{6}$Department of Physics and Astronomy, University of Victoria,
  Victoria, BC V8P 5C2, Canada}

\begin{document}
  
\date{Accepted 2010 March 24. Received 2010 Feb 17; in original form 2009 October 30}

\pagerange{\pageref{firstpage}--\pageref{lastpage}} \pubyear{2009}

\maketitle

\label{firstpage}
\begin{abstract} 

We present six simulations of Galactic stellar haloes formed by the
tidal disruption of accreted dwarf galaxies in a fully cosmological
setting. Our model is based on the Aquarius project, a suite of high
resolution N-body simulations of individual dark matter haloes. We tag
subsets of particles in these simulations with stellar populations
predicted by the \galform{} semi-analytic model. Our method
self-consistently tracks the dynamical evolution and disruption of
satellites from high redshift. The luminosity function and structural
properties of surviving satellites, which agree well with
observations, suggest that this technique is appropriate. We find that
accreted stellar haloes are assembled between $1<z<7$ from less than 5
significant progenitors. These progenitors are old, metal-rich
satellites with stellar masses similar to the brightest Milky Way
dwarf spheroidals ($10^{7}-10^{8}\;\rm{M_{\sun}}$). In contrast to
previous stellar halo simulations, we find that several of these major
contributors survive as self-bound systems to the present day. Both
the number of these significant progenitors and their infall times are
inherently stochastic. This results in great diversity among our
stellar haloes, which amplifies small differences between the
formation histories of their dark halo hosts. The masses
($\sim10^{8}-10^{9}\;\rm{M_{\sun}}$) and density/surface-brightness
profiles of the stellar haloes (from 10--100~kpc) are consistent with
expectations from the Milky Way and M31. Each halo has a complex
structure, consisting of well-mixed components, tidal streams, shells
and other subcomponents. This structure is not adequately described by
smooth models. The central regions ($<10$~kpc) of our haloes are
highly prolate ($c/a\sim0.3$), although we find one example of a
massive accreted thick disc. Metallicity gradients in our haloes are
typically significant only where the halo is built from a small number
of satellites. We contrast the ages and metallicities of halo stars
with surviving satellites, finding broad agreement with recent
observations.

\end{abstract}


\section{Introduction}
\label{sec:introduction}

An extended and diffuse stellar halo envelops the Milky Way. Although
only an extremely small fraction of the stars in the Solar
neighbourhood belong to this halo, they can be easily recognized by
their extreme kinematics and metallicities. Stellar populations with
these properties can now be followed to distances in excess of
100~kpc using luminous tracers such as RR Lyraes, blue horizontal
branch stars, metal-poor giants and globular clusters
(e.g. Oort 1926; Baade 1944; Eggen, Lynden-Bell \& Sandage 1962; Searle \& Zinn 1978; Yanny {et~al.} 2000; Vivas \& Zinn 2006; Morrison {et~al.} 2009).

In recent years, large samples of halo-star velocities
(e.g. Morrison {et~al.} 2000; Starkenburg {et~al.} 2009) \newt{and}
`tomographic' photometric and spectroscopic surveys have shown that
the stellar halo is not a single smoothly-distributed entity, but
instead a superposition of many components
(Belokurov {et~al.} 2006; Juri{\'c} {et~al.} 2008; Bell {et~al.} 2008; Carollo {et~al.} 2007, 2009; Yanny {et~al.} 2009). Notable
substructures in the Milky Way halo include the broad stream of stars
from the disrupting Sagittarius dwarf galaxy
(Ibata, Gilmore \& Irwin 1994; Ibata {et~al.} 2001), extensive and diffuse overdensities
(Juri{\'c} {et~al.} 2008; Belokurov {et~al.} 2007a; Watkins {et~al.} 2009), the Monoceros
`ring' (Newberg {et~al.} 2002; Ibata {et~al.} 2003; Yanny {et~al.} 2003), the orphan stream
(Belokurov {et~al.} 2007b) and other kinematically cold debris
(Schlaufman {et~al.} 2009). Many of these features remain unclear. At
least two kinematically distinct `smooth' halo components have been
identified from the motions of stars in the Solar neighbourhood, in
addition to one or more `thick disc' components
(Carollo {et~al.} 2009). Although current observations only hint at
the gross properties of the halo and its substructures, some general
properties are well-established: the halo is extensive
($>100\:\rm{kpc}$), metal-poor
($\mathrm{\langle[Fe/H]\rangle\sim{-1.6}}$,
e.g. Laird {et~al.} 1988; Carollo {et~al.} 2009) and contains of the order
of 0.1-1\% of the total stellar mass of the Milky Way
\newt{(recent reviews
    include Freeman \& Bland-Hawthorn 2002; Helmi 2008)}.

Low surface-brightness features seen in projection around other galaxies
aid in the interpretation of the Milky Way's stellar halo, and vice
versa. Diffuse concentric `shells' of stars on $100$~kpc scales around
otherwise regular elliptical galaxies \newt{have been} attributed to
accretion events (e.g. Schweizer 1980; Quinn 1984). Recent surveys of
M31 (e.g. Ferguson {et~al.} 2002; Kalirai {et~al.} 2006; Ibata {et~al.} 2007; McConnachie {et~al.} 2009) have
revealed an extensive halo (to $\sim150\,\rm{kpc}$) also displaying
abundant substructure. The surroundings of other nearby Milky Way
analogues are now being targeted by observations using
resolved star counts to reach very low effective surface brightness
limits, although as yet no systematic survey has been carried out to
sufficient depth.
(e.g. Zibetti \& Ferguson 2004; McConnachie {et~al.} 2006; de~Jong, Radburn-Smith \&  Sick 2008; Barker {et~al.} 2009; Ibata, Mouhcine, \& Rejkuba 2009).
A handful of deep observations beyond the Local Group suggest that
stellar haloes are ubiquitous and diverse
(e.g. Sackett {et~al.} 1994; Shang {et~al.} 1998; Malin \& Hadley 1999; Mart{\'{i}}nez-Delgado {et~al.} 2008, 2009; Fa{\'{u}}ndez-Abans {et~al.} 2009).

Stellar haloes formed from the debris of disrupted satellites are a
natural byproduct of hierarchical galaxy formation in the \lcdm{}
cosmology\footnote{In addition to forming components of the accreted
  stellar halo, infalling satellites may cause dynamical heating of a
  thin disc formed `in situ'
  (e.g. Toth \& Ostriker 1992; Velazquez \& White 1999; Benson {et~al.} 2004; Kazantzidis {et~al.} 2008) and may
  also contribute material to an accreted thick disc (Abadi, Navarro \& Steinmetz 2006)
  or central bulge. We discuss these additional contributions to the
  halo, some of which are not included in our modelling, in
  \mnsec{sec:defining_haloes}}. The entire assembly history of a
galaxy may be encoded in the kinematics, metallicities, ages and
spatial distributions of its halo stars. Even though these stars
constitute a very small fraction of the total stellar mass, the
prospects are good for recovering such information from the haloes of
the Milky Way, M31 and even galaxies beyond the Local Group
(e.g. Johnston, Hernquist \& Bolte 1996; Helmi \& White 1999). In this context, theoretical
models can provide useful `blueprints' for interpreting the great
diversity of stellar haloes and their various sub-components, and for
relating these components to fundamental properties of galaxy
formation models. Alongside idealised models of tidal disruption,
\textit{ab initio} stellar halo simulations in realistic cosmological
settings are essential for \newt{direct} comparison with observational
data.

In principle, hydrodynamical simulations are well-suited to this task,
as they incorporate the dynamics of a baryonic component
self-consistently. However, many uncertainties remain in how physical
processes such as star formation and supernova feedback, which act
below the scale of individual particles or cells, are implemented in
these simulations. The computational cost of a single state-of-the-art
hydrodynamical simulation is extremely high. This cost severely limits
the number of simulations that can be performed, restricting the
freedom to explore different parameter choices or alternative
assumptions within a particular model. The computational demands of
hydrodynamical simulations are compounded in the case of stellar halo
models, in which the stars of interest constitute only $\sim1\%$ of
the total stellar mass of a Milky Way-like galaxy. Even resolving the
accreted dwarf satellites in which a significant proportion of these
halo stars may originate is close to the limit of current simulations
of disc galaxy formation. To date, few hydrodynamical simulations have
focused explicitly on the accreted stellar halo (recent examples
include Bekki \& Chiba 2001; Brook {et~al.} 2004; Abadi {et~al.} 2006 and
Zolotov {et~al.} 2009).

In the wider context of simulating the `universal' population of
galaxies in representative ($\gtrsim100\,\rm{Mpc^{3}}$) cosmological
volumes, these practical limitations of hydrodynamical simulations have
motivated the development of a powerful and highly successful
alternative, which combines two distinct modelling techniques:
well-understood high-resolution N-body simulations of large-scale
structure evolved self-consistently from \lcdm{} initial conditions and
fast, adaptable semi-analytic models of galaxy formation with very low
computational cost per run
(Kauffmann, Nusser \&  Steinmetz 1997; Kauffmann {et~al.} 1999; Springel {et~al.} 2001; Hatton {et~al.} 2003; Kang {et~al.} 2005; Springel {et~al.} 2005; Bower {et~al.} 2006; Croton {et~al.} 2006; De~Lucia {et~al.} 2006). In
this paper we describe a technique motivated by this approach which
exploits computationally expensive, ultra-high-resolution N-body
simulations of \textit{individual} dark matter haloes by combining them
with semi-analytic models of galaxy formation. Since our aim is to study
the spatial and kinematic properties of stellar haloes formed through
the tidal disruption of satellite galaxies, our technique goes beyond
standard semi-analytic treatments.

The key feature of the method presented here is the dynamical
association of stellar populations (predicted by the semi-analytic
component of the model) with sets of \textit{individual particles} in
the N-body component. We will refer to this technique as `particle
tagging'. We show how it can be applied by combining the Aquarius suite
of six \newt{high resolution} isolated $\sim10^{12}\,\rm{M_{\sun}}$ dark
matter haloes (Springel {et~al.} 2008a,b) with the
\galform{} semi-analytic model (Cole {et~al.} 1994, 2000; Bower {et~al.} 2006). These simulations \newt{can resolve structures down to
$\sim10^{6}\rm{M_{\sun}}$, comparable to the least massive dark halo
hosts inferred for Milky Way satellites
(e.g. Strigari {et~al.} 2007; Walker {et~al.} 2009)}.

Previous implementations of \newt{the particle-tagging approach}
\movt{(White \& Springel 2000; Diemand, Madau \& Moore 2005; Moore {et~al.} 2006; Bullock \& Johnston 2005; De~Lucia \& Helmi 2008)} have so
far \newt{relied on cosmological simulations} severely limited by
resolution \newt{(Diemand {et~al.} 2005; De~Lucia \& Helmi 2008) or \newt{else} simplified
\newt{higher resolution} N-body models (Bullock \& Johnston 2005)}. \movt{In the
present paper, \newt{we apply this technique} as a postprocessing
operation to a `fully cosmological' simulation, in which structures have
grown \textit{ab initio}, interacting with one another
self-consistently. \newt{The resolution of our simulations is sufficient
to resolve stellar halo substructure in considerable detail.}}

With the aim of presenting our modelling approach and exploring some
of the principal features of our simulated stellar haloes, we proceed
as follows. In \mnsec{sec:method_overview} we review the Aquarius
simulations and their post-processing with the \galform{} model, and
in \mnsec{sec:buildinghaloes} we describe our method for recovering
the spatial distribution of stellar populations in the halo by tagging
particles. \newt{We calibrate our model by comparing the statistical
  properties of the surviving satellite population to observations;
  the focus of this paper is on the stellar halo, rather on than the
  properties of these satellites.} In \mnsec{sec:results} we
describe our model stellar haloes and compare their structural
properties to observations of the Milky Way and M31. We also examine
the assembly history of the stellar haloes in detail
(\mnsec{sec:assembly}) and explore the relationship between the haloes
and the surviving satellite population. Finally, we summarise our
results in \mnsec{sec:conclusion}.

\section{Aquarius and Galform}
\label{sec:method_overview}

Our model has two key components: the Aquarius suite of six
high-resolution N-body simulations of Milky Way-like dark matter haloes,
and \galform{}, a semi-analytic model of galaxy formation. The technique
of post-processing an N-body simulation with a semi-analytic model is
well established
(Kauffmann {et~al.} 1999; Springel {et~al.} 2001; Helly {et~al.} 2003; Hatton {et~al.} 2003; Kang {et~al.} 2005; Bower {et~al.} 2006; De~Lucia {et~al.} 2006),
although its application to high-resolution simulations of individual
haloes such as Aquarius is novel and we review relevant aspects of the
\galform{} code in this context below.

\newt{Here, i}n the post-processing of the N-body simulation,
the stellar populations predicted by \galform{} to form in each halo
are \newt{also} associated with `tagged' subsets of dark matter
particles. By following these tagged dark matter particles, we track
the evolving \textit{spatial distribution and kinematics} of their associated
stars, in particular those that are stripped from satellites to build
the stellar halo. This level of detail regarding the distribution of
halo stars is unavailable to a standard semi-analytic approach, in
which the structure of each galaxy is represented by a combination of
analytic density profiles.

\newt{Tagging particles in this way
  requires} the fundamental assumption that baryonic mass nowhere
dominates the potential and hence does not perturb the collisionless
dynamics of the dark matter. Generally, a massive thin disc is
expected to form at some point in the history of our \newt{`main'}
haloes. Although our semi-analytic model accounts for this thin disc
consistently, our dark matter tagging scheme cannot represent its
dynamics. For this reason, and also to avoid
confusion with our accreted halo stars, we do
not attempt to tag dark matter to represent stars forming in situ in a
thin disc at the centre of the main halo. \newt{T}he approximation \newt{that} the dynamics
of stars can be fairly represented by tagging dark matter
particles is justifiable for systems with high mass-to-light
ratios such as the dwarf satellites of the Milky Way and M31
\movt{(e.g. Simon \& Geha 2007; Walker {et~al.} 2009)}, the units from which
stellar haloes are assembled in our models.

\subsection{The Aquarius Haloes}
\label{sec:aquarius}

Aquarius (Springel {et~al.} 2008a) is a suite of high-resolution
simulations of six dark matter haloes having masses within the range
$1-2\times10^{12}\,\rm{M_{\sun}}$, comparable to values typically
inferred for the Milky Way halo
(e.g. Battaglia {et~al.} 2005; Smith {et~al.} 2007; Li \& White 2008; Xue {et~al.} 2008). By matching the
abundance of dark matter haloes in the Millennium simulation to the SDSS
stellar mass function, Guo {et~al.} (2009) find
$2.0\times10^{12}\,\rm{M_{\sun}}$ (with a 10-90\% range of
$0.8\times10^{12}\rm{M_{\sun}}$ to
$4.7\times10^{12}\rm{M_{\sun}}$). This value is sensitive to the
assumption that the Milky Way is a typical galaxy, and to the adopted
Milky Way stellar mass ($5.5\times10^{10}\,\rm{M_{\sun}}$;
Flynn {et~al.} 2006).

\newt{The Aquarius haloes were selected from a lower resolution version of the
    Millennium-II simulation (Boylan-Kolchin {et~al.} 2009) and individually
  resimulated using a multi-mass particle (`zoom') technique. In this
  paper we use the `level 2' Aquarius simulations, the highest level
  at which all six haloes were simulated.} We refer the reader to
Springel {et~al.} (2008a,b) for a comprehensive account
of the entire simulation suite and demonstrations of numerical
convergence. We list relevant properties of each halo/simulation in
\mntab{tbl:aquarius}.  The simulations were carried out with the
parallel Tree-PM code \gadgetthree{}, an updated version of
\gadgettwo{} (Springel 2005). The Aq-2 simulations used a
fixed comoving Plummer-equivalent gravitational softening length of
$\epsilon = 48\:h^{{-}1} \:\rm{pc}$. \lcdm{} cosmological parameters
were adopted as $\Omega_{\rm{m}} = 0.25$, $\Omega_{\Lambda} = 0.75$,
$\sigma_{8} = 0.9$, $n_{\rm{s}} = 1$, and Hubble constant $H_{0} =
100h \,\rm{km\,s}^{{-}1}\rm{Mpc}^{-1}$. A value of $h = 0.73$ is
assumed throughout this paper. These parameters are identical to those
used in the Millennium Simulation and are marginally
consistent with WMAP 1- and 5-year constraints
(Spergel {et~al.} 2003; Komatsu {et~al.} 2009).

\begin{table}
    \caption{Properties of the six Aquarius dark matter halo
      simulations (Springel {et~al.} 2008a) on which the models in this
      paper are based. The first column labels the simulation
      (abbreviated from the as Aq-A-2, Aq-B-2 etc.). From left to
      right, the remaining columns give the particle mass
      $m_{\rm{p}}$, the number of particles within $r_{200}$, the
      virial radius at $z=0$; the virial mass of the halo, $M_{200}$;
      and the maximum circular velocity, $V_{max}$, and corresponding
      radius, $r_{\rm{max}}$. Virial radii are defined as the radius
      of a sphere with mean inner density equal to 200 times the
      critical density for closure.}
    \begin{tabular}{@{}lcccccc}
\hline
 & $m_{\rm{p}}$          & $N_{200}$   & $M_{200}$         & $r_{200}$    & $V_{\rm{max}}$      & $r_{\rm{max}}$ \\
 & $[10^{3}\rm{M_{\sun}}]$ & $[10^{6}]$ & $[10^{12}\rm{M_{\sun}}]$ & $[\rm{kpc}]$ & $[\rm{km}\,\rm{s}^{{-1}}]$ & $[\rm{kpc}]$ \\
\hline
  A &$13.70$ & 135 & $1.84$& 246 & 209 & 28  \\
  B &$6.447$ & 127 & $0.82$& 188 & 158 & 40  \\
  C &$13.99$ & 127 & $1.77$& 243 & 222 & 33  \\
  D &$13.97$ & 127 & $1.74$& 243 & 203 & 54  \\
  E &$9.593$ & 124 & $1.19$& 212 & 179 & 56  \\
  F &$6.776$ & 167 & $1.14$& 209 & 169 & 43  \\
\hline
    \end{tabular}
 \label{tbl:aquarius}
\end{table}

\subsection{The \galform{} Model}
\label{sec:galform}

N-body simulations of cosmic structure formation supply information on
the growth of dark matter haloes, which can serve as the starting
point for a semi-analytic treatment of baryon accretion, cooling and
star formation (see Baugh 2006, for a comprehensive discussion of the
  fundamental principles of semi-analytic
  modelling). The Durham semi-analytic model,
\galform{}, is used in this paper to postprocess the Aquarius N-body
simulations. \movt{The \galform{} code is controlled by a number of
  interdependent parameters which are constrained in part by
  theoretical limits and results from hydrodynamical
  simulations. Remaining parameter values are chosen such that the
  model satisfies statistical comparisons with several datasets, for
  example the galaxy luminosity function measured in several wavebands
  (e.g. Baugh {et~al.} 2005; Bower {et~al.} 2006; Font {et~al.} 2008). Such statistical constraints
  on large scales do not guarantee that the same model will provide a
  good description of the evolution of a single `Milky Way' halo and
  its satellites. \movt{A model producing a satellite \newt{galaxy}
    luminosity function consistent with observations is a fundamental
    prerequisite for the work presented here, in which a proportion of
    the total satellite population provides the raw material for the
    assembly of stellar haloes.} \newt{W}e demonstrate below that the
  key processes driving galaxy formation on small scales are captured
  to good approximation by the existing \galform{} model and
  parameter values of
  Bower {et~al.} (2006).}

Many of the physical processes of greatest relevance to galaxy
formation on small scales were explored within the context of
semi-analytic modelling by Benson {et~al.} (2002b). Of particular
significance are the suppression of baryon accretion and cooling in
low mass haloes as the result of photoheating by a cosmic ionizing
background, and the effect of supernova feedback
in shallow potential wells. Together, these effects constitute a
straightforward astrophysical explanation for the disparity between
the number of low mass dark subhaloes found in N-body simulations of
Milky Way-mass hosts and the far smaller number of luminous satellites
observed around the Milky Way (the so-called `missing satellite'
problem). Recent discoveries of faint dwarf satellites and an improved
understanding of the completeness of the Milky Way sample (Koposov {et~al.} 2008; Tollerud {et~al.} 2008, and
  refs. therein) have reduced the deficit of
\textit{observed} satellites, to the point of qualitative agreement
with the prediction of the model of
Benson {et~al.} (2002b). \newt{A}t issue now is the
quality (rather than the lack) of agreement between such models and
the data. \movt{\newt{W}e pay particular attention to the suppressive
  effect of photoheating. This is a significant process for shaping
  the faint end of the satellite luminosity function when, as we
  assume here, the strength of supernova feedback is fixed by
  constraints on the galaxy population as a whole.}

\subsubsection{Reionization and the satellite luminosity function}

A simple model of reionization heating based on a halo mass dependent
cooling threshold (Benson {et~al.} 2003) is implemented in the
Bower {et~al.} (2006) model of \galform{}. This threshold is set by
parameters termed $V_{\rm{cut}}$ and $z_{\rm{cut}}$. No gas is
allowed to cool within haloes having a circular velocity below
$V_{\rm{cut}}$ at redshifts below $z_{\rm{cut}}$. To good
approximation, this scheme reproduces the link between the suppression
of cooling and the evolution of the `filtering mass' (as defined
  by Gnedin 2000) found in the more detailed model of
Benson {et~al.} (2002b), where photoheating of the intergalactic medium
was modelled explicitly. In practice, in this simple model, the value
of $V_{\rm{cut}}$ is most important. Variations in $z_{\rm{cut}}$
within plausible bounds have a less significant effect on the $z=0$
luminosity function.

As stated above, we adopt as a fiducial model the \galform{}
implementation and parameters of Bower {et~al.} (2006). However, we make a
single parameter change, lowering the value of $V_{\rm{cut}}$ from
$50\,\rm{km\,s^{-1}}$ to $30\,{\rm{km\,s^{-1}}}$. This choice is
motivated by recent \textit{ab initio} cosmological galaxy formation
simulations incorporating the effects of photoionization
self-consistently (Hoeft {et~al.} 2006; Okamoto, Gao \& Theuns 2008; Okamoto \& Frenk 2009; Okamoto {et~al.} 2009). These studies find that values of
$V_{\rm{cut}}\sim25-35\,\rm{km\,s^{-1}}$ are preferable to the higher
value suggested by the results of Gnedin (2000) and adopted in
previous semi-analytic models
(e.g. Somerville 2002; Bower {et~al.} 2006; Croton {et~al.} 2006; Li, De~Lucia \&  Helmi 2009a). \newt{Altering
  this value affects only the very faint end of the galaxy luminosity
  function, and so does not change the results of (Bower {et~al.} 2006).}
\movt{The choice of a fiducial set of semi-analytic parameters in this
  paper \newt{illustrates} the flexibility
  \newt{of} our approach to modelling stellar
  haloes. The N-body component of our models -- Aquarius -- represents
  a considerable investment of computational time. In contrast, the
  semi-analytic post-processing can be re-run in only a few hours, and
  can be easily `upgraded' (by adding physical processes and
  constraints) in order to provide more detailed output, explore the
  consequences of parameter variations, or compare alternative
  semi-analytic models.}

\begin{figure}
\includegraphics[width=84mm]{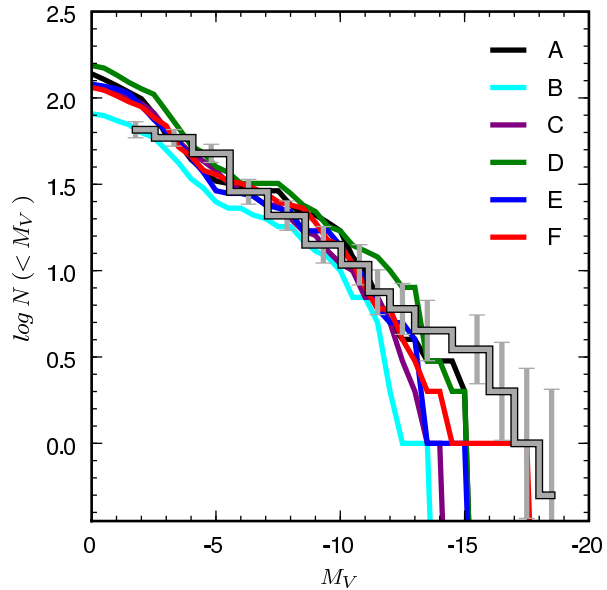}
\caption{The cumulative \textit{V}-band luminosity functions (LFs) of
  satellite galaxies for the six Aquarius haloes, adopting in
  \galform{} the parameters of Bower {et~al.} (2006) with $V_{\rm{cut}} =
  30\:\rm{km\,s^{-1}}$. These LFs \textit{include} the effects of
  tidal stripping measured from our assignment of stars to dark matter
  particles (\mnsec{sec:buildinghaloes}), although this makes only a
  small difference to the LF from our semi-analytic model alone. All
  galaxies within 280~kpc of the halo centre are counted as satellites
  (the total number of contributing satellites in each halo is
  indicated in the legend). The stepped line (grey, with error bars)
  shows the observed mean luminosity function found by
  Koposov {et~al.} (2008) for the MW and M31 satellite system (also to
  280~kpc), assuming an NFW distribution for satellites in correcting
  for SDSS sky coverage and detection efficiency below
  $M_{\mathrm{v}}=-10$. The colour-coding of our haloes in this figure
  is used throughout.}
\label{fig:galform_lf}
\end{figure}

The \textit{V}-band satellite luminosity function resulting from the
application of the \galform{} model described above to each Aquarius
halo is shown in \fig{fig:galform_lf}. Satellites are defined as all
galaxies within a radius of 280 kpc from the centre of potential in
the principal halo, equivalent to the limiting distance of the
Koposov {et~al.} (2008) completeness-corrected observational sample. These
luminosity functions are measured from the \textit{particle}
realisations of satellites that we describe in the following section,
and not directly from the semi-analytic model. They therefore account
for the effects of tidal stripping, although these are minor: the
fraction of satellites brighter than $M_{\rm{V}}=-10$ is reduced very
slightly in some of the haloes. In agreement with the findings of
Benson {et~al.} (2002a), the model matches the faint end of the
luminosity function well, but fewer bright satellites are found in
each of our six models than are observed in the mean of the Milky Way
+ M31 system, although the number of objects concerned is small. The
true abundance of bright satellites for Milky Way-mass hosts is poorly
constrained at present, so it is unclear whether or not this
discrepancy reflects cosmic variance, a disparity in mass between the
Aquarius haloes and the Milky Way halo, or a shortcoming of our
fiducial Bower {et~al.} (2006) model. A modification of this model in which
the tidal stripping of hot gas coronae around infalling satellites is
explicitly calculated (rather than assuming instantaneous
  removal; see Font {et~al.} 2008) produces an acceptable abundance of bright
satellites.

\subsubsection{Further details}

Within \galform{}, cold gas is transferred from tidally destroyed 
satellites to the disc of the central galaxy when their host subhaloes
are no longer identified at the resolution limit imposed by
\subfind{}. In the Aq-2 simulations this corresponds to a minimum
resolved dark halo mass of $\sim3\times10^{5}\rm{M}_{\sun}$. In the
\galform{} model of Bower {et~al.} (2006), which does not include tidal
stripping or a `stellar halo' component, the satellite galaxy is
considered to be fully disrupted (merged) at this point: its stars
are transferred to the bulge component of the central galaxy. By
contrast, our particle representation (described in
\mnsec{sec:buildinghaloes}) allows us to follow the \textit{actual}
fate of the satellite stars independently of this choice in the
semi-analytic model. This choice is therefore largely a matter of
`book-keeping'; we have ensured that adopting this approach does not
prematurely merge galaxies in the semi-analytic model that are still
capable of seeding new stellar populations into the particle
representation. Semi-analytic models based on N-body simulations
often choose to `follow' satellites with dark haloes falling below
the numerical resolution by calculating an appropriate merger
time-scale from the last-known N-body orbital parameters, accounting
for dynamical friction. However, the resolution of Aquarius is
sufficiently high to make a simpler and more self-consistent
approach preferable in this case, preserving the one-to-one
correspondence between star-forming semi-analytic galaxies and bound
objects in the simulation. We have checked that allowing
semi-analytic galaxies to survive without resolved subhaloes,
subject to the treatment of dynamical friction used by
Bower {et~al.} (2006), affects only the faintest ($M_{\mathrm{v}} \sim 0$)
part of the survivor luminosity function. The true nature and
survival of these extremely faint sub-resolution galaxies remains an
interesting issue to be addressed by future semi-analytic models of
galactic satellites.

In \mntab{tbl:summary} (\mnsec{sec:results}) we list the \textit{V}-band
magnitudes and total stellar masses of the central galaxies that form in
the six Aquarius haloes. A wide range is evident, from an M31-analogue
in halo Aq-C, to an M33-analogue in Aq-E. This is not unexpected: the
Aquarius dark haloes were selected only on their mass and isolation, and
these criteria alone do not guarantee that they will host close
analogues of the Milky Way. The scaling and scatter in the predicted
relationship between halo mass and central galaxy stellar mass are
model-dependent. With the \galform{} parameter values of
Bower {et~al.}, the mean central stellar mass in a typical
Aquarius halo ($M_{\rm{halo}}\sim1.4\times10^{12}\mathrm{M_{\sun}}$) is
$\sim1.5\times10^{10}\,\mathrm{M_{\sun}}$, approximately a factor of
3--4 below typical estimates of the stellar mass of the Milky Way
($\sim6\times10^{10}\,\mathrm{M_{\sun}}$ Flynn {et~al.} 2006); the
scatter in $M_{\rm{gal}}$ for our central galaxies reflects the overall
distribution produced by the model of Bower {et~al.} (2006) for haloes of this
mass. The model of De~Lucia {et~al.} (2006), which like the Bower {et~al.} (2006)
model was constrained using statistical properties of bright field and
cluster populations, produces a mean central stellar mass of
$\sim4\times10^{10}\,\mathrm{M_{\sun}}$ for the typical halo mass of the
Aquarius simulations, as well as a smaller scatter about the mean value.

In light of these modelling uncertainties and observational
uncertainties in the determination of the true Milky Way dark halo
mass to this precision, we choose not to scale the Aquarius haloes to
a specific mass for `direct' comparison with the Milky Way. The
results we present concerning the assembly and structure of stellar
haloes and the ensemble properties of satellite systems should not be
sensitive to whether or not their galaxies are predicted to be direct
analogues of the Milky Way by the Bower {et~al.} (2006) \galform{}
model. Therefore, in interpreting the \textit{absolute}
\newt{values} of quantities compared to observational
data in the following sections, it should be borne in mind that we
model a \textit{range} of halo masses that could lie somewhat below
the likely Milky Way value.

The Bower {et~al.} (2006) implementation of \galform{} results in a
mass-metallicity relation for faint galaxies which is slightly steeper
than that derived from the satellites of the Milky Way and M31
(e.g. Mateo 1998; Kirby {et~al.} 2008; see also Tremonti {et~al.} 2004 and
refs. therein). This results in model galaxies being on average
$\sim0.5\,\rm{dex}$ more metal-poor in \feh{} than the observed
relation at magnitudes fainter than $M_{\rm{V}}\sim-10$. Whilst it
would be straightforward to make \textit{ad hoc} adjustments to the
model parameters in order to match this relation, doing so would
violate the agreement established between the Bower {et~al.} (2006)
parameter set and a wide range of statistical constraints from the
bright ($M_V<-19$) galaxy population.

\section{Building Stellar Haloes}

\label{sec:buildinghaloes}

\subsection{Assigning Stars To Dark Matter}
\label{sec:starstodm}

Observations of the stellar velocity distributions of dwarf spheroidal
satellites of the Milky Way imply that these objects are
dispersion-supported systems with extremely high mass-to-light ratios,
of order 10--1000
(e.g. Mateo 1998; Simon \& Geha 2007; Strigari {et~al.} 2007; Wolf {et~al.} 2009; Walker {et~al.} 2009). As
we describe in this section, in order to construct basic models of
these high-M/L systems without simulating their baryon dynamics
explicitly, we will assume that their stars are formed `dynamically
coupled' to a strongly bound fraction of their dominant dark matter
component, and will continue to trace that component throughout the
simulation. Here we further assume that the depth at which stars form
in a halo potential well depends only on the total mass of the
halo. While these assumptions are too simplistic a description of
stellar dynamics in such systems to compare with detailed structural
and kinematic observations, we show that they none the less result in
half-light radii and line-of-sight velocity dispersions in agreement
with those of Milky Way dwarf spheroidals. Hence the disruption of a
fraction of these model satellites by tidal forces in the main halo
should reproduce stellar halo components (`streams') at a level of
detail sufficient for an investigation of the assembly and gross
structure of stellar haloes. \reft{We stress that these comparisons
  are used as constraints on the single additional free parameter in
  our model, and are not intended as predictions of a model
  for the satellite population.}

In the context of our \galform{} model, the stellar content of a
single galaxy can be thought of as a superposition of many distinct
stellar populations, each defined by a particular formation time and
metallicity. Although the halo merger tree used as input to \galform{}
is discretized by the finite number of simulation outputs (snapshots),
much finer interpolating timesteps are taken between snapshots when
solving the differential equations governing star
formation. Consequently, a large number of distinct populations are
`resolved' by \galform{}. However, we can update our particle
(dynamical) data (and hence, can assign stars to dark matter) only at
output times of the pre-existing N-body simulation. For the purposes
of performing star-to-dark-matter assignments we reduce the
fine-grained information computed by \galform{} between one output
time and the next to a single aggregated population of `new stars'
formed at each snapshot.

As discussed above and in \mnsec{sec:introduction}, we adopt the
fundamental assumption that the motions of stars can be represented by
dark matter particles. The aim of our method here is to select a
sample of representative particles from the parent N-body simulation
to trace \textit{each such stellar population}, individually. We
describe first the general objective of our selection process, and
then examine the selection criteria that we apply in practice.

Consider first the case of a single galaxy evolving in isolation. At a
given simulation snapshot (B) the total mass of new stars formed since
the previous snapshot (A) is given by the difference in the stellar
mass of the semi-analytic galaxy recorded at each time,
\begin{equation}
{\Delta}M_{\star}^{AB} = M_{\star}^{B} -M_{\star}^{A}.
\end{equation}
In our terminology, $\Delta M_{\star}^{AB}$ is a single stellar
population (we do not track the small amount of mass lost during
subsequent stellar evolution). The total mass in metals within the
population is determined in the same way as the stellar mass; we do
not follow individual chemical elements. In a similar manner, the
luminosity of the new population (at $z=0$) is given by the difference
of the total luminosities (after evolution to $z=0$) at successive
snapshots.

From the list of particles in the simulation identified with the dark
matter halo of the galaxy at B, we select a subset to be tracers of
the stellar population ${\Delta}M_{\star}^{AB}$. Particles in this
tracer set are `tagged', i.e. are identified with data describing the
stellar population. In the scheme we adopt here, equal `weight'
(fraction of stellar mass, luminosity and metals in $\Delta
M_{\star}^{AB}$) is given to each particle in the set of tracers. We
repeat this process for all snapshots, applying the energy criterion
described below to \textit{select a new set of DM tracers each time
  new stars are formed} in a particular galaxy. In this scheme, the
same DM particle can be selected as a tracer at more than one output
time (i.e. the same DM particle can be tagged with more than one
stellar population). Hence a given DM particle accumulates its own
individual star formation history. The dynamical evolution of
satellite haloes determines whether or not a particular particle is
eligible for the assignment of new stars during any given episode of
star formation.

So far we have considered an `isolated' galaxy. In practice, we apply
this technique to a merger tree, in which a galaxy grows by the
accretion of satellites as well as by \textit{in situ} star
formation. In the expression above, the total stellar mass at A,
$M_{\star}^{A}$, is simply modified to include a sum over $N$
immediate progenitor galaxies in addition to the galaxy itself i.e.,
\begin{equation}
  {\Delta}M_{\star}^{AB} = M_{\star}^{B} - M_{\star,0}^{A} - \sum_{i>0}
  {M_{\star,i}^{A}}
\end{equation}
where $M_{\star,0}^{A}$ represents the galaxy itself and
$M_{\star,i}^{A}$ is the total stellar mass (at A) of the $i$'th
progenitor deemed to have merged with the galaxy in the interval
AB. Stars forming in the progenitors during the interval AB and stars
forming in the galaxy itself are treated as a single population.

There is a one-to-one correspondence between a galaxy and a dark
matter structure (halo or subhalo) from which particles are chosen as
tracers of its newly formed stars. As discussed in
\mnsec{sec:galform}, a satellite galaxy whose host subhalo is no
longer identified by \subfind{} has its cold gas content transferred
immediately to the central galaxy of their common parent halo and
forms no new stars. In the semi-analytic model, the stars of the
satellite are also added to the bulge component of the central
galaxy. This choice is irrelevant in our particle representation, as
we can track the actual fate of these stars.

\begin{figure*}
\centering 
\includegraphics[clip=False]{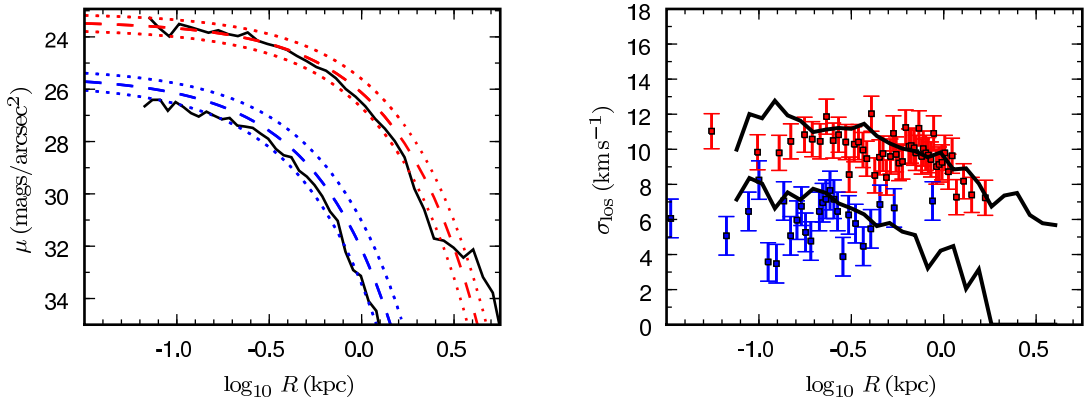}

\caption{Examples of individual satellites in our models (solid black
  lines), compared to Fornax (red) and Carina (blue), showing surface
  brightness (left, Irwin \& Hatzidimitriou 1995) and line-of-sight velocity
  dispersion (right, Walker {et~al.} 2009). With our fiducial
  \galform{} model, simultaneous matches to both $\sigma(R)$ and
  $\mu(R)$ for these datasets are found only among satellites that
  have undergone substantial tidal stripping (see text).}
\label{fig:examplesatA11z0}
\end{figure*}

\subsection{Assignment criteria}
\label{sec:assignmentcriteria}

\subsubsection{Selection of dark matter particles}
\label{sec:method}

In this section we describe how we choose the dark matter particles
within haloes that are to be tagged with a newly formed stellar
population. In \mnsec{sec:introduction} we briefly described the
particle-tagging method employed by Bullock \& Johnston (2005), the philosophy
of which we term \textit{`in vitro'}, using idealised initial
conditions to simulate accretion events individually in a `controlled'
environment. By contrast, our approach is to \textit{postprocess}
fully cosmological simulations \textit{`in vivo'}\footnote{This
  terminology should not be taken to imply that `star particles'
  themselves are included in the N-body simulation; here stellar
  populations are simply tags affixed to dark matter particles.}. In a
fully cosmological N-body simulation the growth of the central
potential, the structure of the halo and the orbits, accretion times
and tidal disruption of subhaloes are fully consistent with one
another. The central potential is non-spherical (although no disc
component is included in our dynamical model) and can grow violently
as well as through smooth accretion. Our model is therefore applicable
at high redshift when the halo is undergoing rapid assembly. The
complexities in the halo potential realised in a fully cosmological
simulation are likely to be an important influence on the dynamics of
satellites (e.g. Sales {et~al.} 2007a) and on the evolution of
streams, through phase-mixing and orbital precession
(e.g. Helmi \& White 1999).

\newt{W}e approach the selection of dark matter particles for stellar
tagging differently to Bullock \& Johnston (2005)\newt{, because}\movt{ we are
postprocessing a cosmological N-body simulation rather than constructing
idealised initial conditions for each satellite}. \newt{R}ather than
assigning the mass-to-light ratio of each tagged particle by comparing
stellar and dark matter energy distribution functions in the halo
concerned, we assume that the energy distribution of newly formed stars
traces that of the dark matter. We order the particles in the
halo by binding energy\footnote{Here, the most bound particle is that
with the most negative total energy, including both kinetic and
gravitational contributions. Binding energies are computed relative to
the bound set of particles comprising an object identified by
\subfind{}.} and select a most-bound fraction $f_{\rm{MB}}$ to be
tagged with newly-formed stars. As previously described, stars are
shared equally among the selected DM particles.

\newt{Our approach implies a rather simple dynamical model for stars
  in satellite galaxies. However, the main results of this paper do
  not concern the satellites themselves; instead we focus on the
  debris of objects that are totally (or largely) disrupted to build
  the stellar halo. As we describe below, we compare the structure and
  kinematics of our model satellites (those that survive at $z=0$) to
  Local Group dwarf galaxies in order to fix the value of the free
  parameter, $f_{\rm{MB}}$. Since we impose this constraint, our
  method cannot predict these satellite properties \textit{ab
    initio}. Constraining our model in this way ensures reasonable
  structural properties in the population of progenitor satellites,
  and retains full predictive power with regard to the stellar
  halo. More complex models would, of course, be possible, in which
  $f_{\rm{MB}}$ is not a free parameter but is instead physically
  determined by the semi-analytic model. It would also be possible to
  use a more complicated tagging scheme to attempt to represent, for
  example, star formation in a disc. However, }
\movt{\newt{such models would add substantial complexity to
    the method and there are currently very few observational
    constraints on how stars were formed in satellite galaxies. Thus,
    we believe that a}  simple model suffices for our present
  study of the stellar halo\newt{.}}


Our approach has similarities with that of De~Lucia \& Helmi (2008), who tag
the most bound 10\% of particles in satellite haloes with
stars. However, De~Lucia \& Helmi perform this tagging only
\textit{once} for each satellite, at the time at which its parent halo
becomes a subhalo of the main halo (which we refer to as the time of
infall\footnote{In both Bullock \& Johnston (2005) and De~Lucia \& Helmi (2008) only
  satellites directly accreted by the main halo `trigger' assignments
  to dark matter; the hierarchy of mergers/accretions forming a
  directly-infalling satellite are subsumed in that single
  assignment.}). Both this approach and that of Bullock \& Johnston (2005)
define the end result of the previous dynamical evolution of an
infalling satellite, the former by assuming light traces dark matter
and the latter with a parameterized King profile.

As described above, in our model each newly-formed stellar population
is assigned to a subset of DM particles, chosen according to the
`instantaneous' dynamical state of its host halo. This choice is
independent of any previous star formation in the same halo. It is the
dynamical evolution of these many tracer sets in each satellite that
determines its stellar distribution at any point in the simulation.

Implementing a particle-tagging scheme such as this within a fully
cosmological simulation requires a number of additional issues to be
addressed, which we summarise here.

\renewcommand{\labelenumii}{\roman{enumi}}
\begin{enumerate}
\item \textit{Subhalo assignments}: Star formation in a satellite
  galaxy will continue to tag particles regardless of the level of
  its halo in the hierarchy of bound structures (halo, subhalo,
  subsubhalo etc.). The growth of a dark matter halo ends when it
  becomes a subhalo of a more massive object, whereupon its mass is
  reduced through tidal stripping. The assignment of stars to
  particles in the central regions according to binding energy
  should, of course, be insensitive to the stripping of dark matter
  at larger radii.  However, choosing a fixed fraction of dark
  matter tracer particles to represent new stellar populations
  couples the mass of the subhalo to the number of particles
  chosen. Therefore, when assigning stars to particles in a subhalo,
  we instead select a fixed \textit{number} of particles, equal to
  the number constituting the most-bound fraction $f_{\rm{MB}}$ of
  the halo at the time of infall.

\item \textit{Equilibrium criterion}: To guard against assigning stars
  to sets of tracer particles that are temporarily far from dynamical
  equilibrium, we adopt the conservative measure of deferring
  assignments to any halo in which the centres of mass and potential
  are separated by more than 7\% of the half-mass radius $r_{1/2}$. We
  select $0.07\,r_{1/2}$ in accordance with the criterion of
  $0.14\,r_{\rm{vir}}$ used to select relaxed objects in the study of
  Neto {et~al.} (2007), taking $r_{\rm{vir}}\sim2\,r_{1/2}$. These deferred
  assignments are carried out at the next snapshot at which this
  criterion is satisfied, or at the time of infall into a more massive
  halo.

\item \textit{No in situ star formation}: Stars formed in the
  main galaxy in each Aquarius simulation (identified as the central
  galaxy of the most massive dark halo at $z=0$) are never assigned to
  DM particles. This exclusion is applied over the entire history of
  that galaxy. Stars formed in situ are likely to contribute to the
  innermost regions of the stellar halo, within which they may be
  redistributed in mergers. However, the dynamics of stars formed in a
  dissipationally-collapsed, baryon-dominated thin disc cannot be
  represented with particles chosen from a dark matter-only
  simulation. We choose instead to study the accreted component in
  isolation. Our technique none the less offers the possibility of
  extracting \textit{some} information on a fraction of in situ stars
  were we to assign them to dark matter particles (those contributing
  to the bulge or forming at early times, for example). We choose to
  omit this additional complexity here. SPH simulations of stellar
  haloes (which naturally model the in situ component more accurately
  than the accreted component) suggest that the contribution of in
  situ stars to the halo is small beyond $\sim20$~kpc
  (Abadi {et~al.} 2006; Zolotov {et~al.} 2009).

  At early times, when the principal halo in each simulation is
  growing rapidly and near-equal-mass mergers are common, the
  definition of the `main' branch of its merger tree can become
  ambiguous. Also, the main branch of the galaxy merger tree need not
  follow the main branch of the halo tree. Hence, our choice of which
  branch to exclude (on the basis that it is forming `in situ' stars)
  also becomes ambiguous; indeed, it is not clear that any of these
  `equivalent' early branches should be excluded. Later we will show
  that two of our haloes have concentrated density profiles. We have
  confirmed that these \textit{do not} arise from making the `wrong'
  choice in these uncertain cases, i.e. from tagging particles in the
  dynamically robust core of the `true' main halo. Making a different
  choice of the excluded branch in these cases (before the principal
  branch can be unambiguously identified) simply replaces one of these
  concentrated components with another very similar
  component. Therefore, we adopt the above definition of the galaxy
  main branch when excluding in situ stars.

\end{enumerate}

\subsubsection{Individual satellites}

We show in \newt{the following section}
that with  a suitable choice of
the most-bound fraction, our method produces \reft{a population of}
model satellites at $z=0$ \newt{having} properties
consistent with observed relationships between
magnitude, half-light radius/surface brightness and velocity
dispersion for satellite populations of the Milky Way and M31. In
\fig{fig:examplesatA11z0} we show profiles of surface brightness and
velocity dispersion for two individual satellites from these models at
$z=0$, chosen to give a rough match to observations of Fornax and
Carina. This suggests that our galaxy formation model and the simple
prescription for the spatial distribution of star formation
\reft{can} produce realistic stellar structures within dark
haloes. However, while it is possible to match these \reft{individual}
observed satellites with examples drawn from our models, we caution
that we can only match their observed surface brightness and velocity
dispersion profiles \textit{simultaneously} by choosing model
satellites that have suffered substantial tidal stripping. \newt{This
  is most notable in the case of} \newt{o}ur match to
Fornax\newt{, which} retains only 2\% of its dark matter relative to
the time of its accretion to the main halo, and 20\% of its stellar
mass. \newt{However, as we show in \mnsec{sec:assembly}, the majority
  of massive surviving satellites have not suffered substantial tidal
  stripping.}

We have tested our method with assignments for each satellite
delayed until the time of infall, as in De~Lucia \& Helmi (2008). This
results in slightly more compact galaxies than in our standard
\textit{in vivo} approach, where mergers and tidal forces (and
relaxation through two-body encounters for objects near the
resolution limit) can increase the energies of tagged dark matter
particles. However, we find that this makes little difference to the
results that we discuss below.

\subsubsection{Parameter constraints and convergence}
\label{sec:constraints}

\begin{figure}
\includegraphics[]{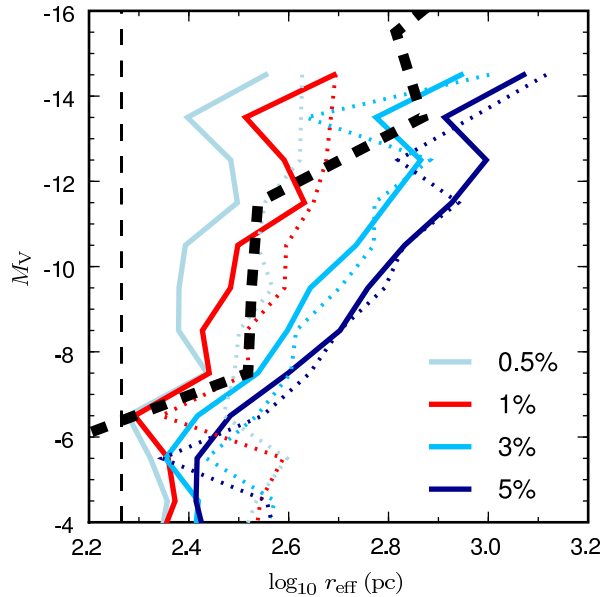}
\caption{Median effective radius \newt{$r_{\rm{eff}}$} (enclosing half
  of the total luminosity in projection) as a function of magnitude
  for model satellites in haloes Aq-A and Aq-F at $z=0$. \newt{A thin
    vertical dashed line indicates the softening scale of the
    simulation: $r_{\rm{eff}}$ is unreliable close to this value and
    meaningless below it.}  \textit{Thick lines} represent \newt{our}
  higher-resolution simulation\newt{s} (Aq-2) using a range of values
  of the fraction of most bound particles chosen in a stellar
  population assignment, $f_{\rm{MB}}$. \textit{Dotted lines}
  correspond to lower resolution simulations (Aq-3) of the same
  haloes. A \textit{thick dashed line} shows the corresponding median
  of \newt{observations of} Local Group dwarf galaxies. These
  galaxies, and our model data points for all haloes in the Aq-2
  series with $f_{\rm{MB}}=1\%$, are plotted individually in
  \fig{fig:satellite_relations}.}
\label{fig:resolutiontests}
\end{figure}

We now compare the $z=0$ satellite populations of our models with
trends observed in the dwarf companions of the Milky Way and M31 in
order to determine a suitable choice of the fixed fraction,
$f_{\rm{MB}}$, of the most bound dark matter particles selected in a
given halo. \newt{O}ur aim is to study the stellar halo, \newt{and
  therefore} we use the sizes of our \newt{surviving} satellites as a
constraint on $f_{\rm{MB}}$ and as a test of convergence. Within the
range of $f_{\rm{MB}}$ that produces plausible satellites, the gross
properties of our haloes, such as total luminosity, change by only a
few percent.

In \fig{fig:resolutiontests}, we show the relationship between the
absolute magnitudes, $M_{\rm{V}}$, of satellites (combining data from
two of our simulations, Aq-A and Aq-F), and the projected radius
enclosing one half of their total luminosity, which we refer to as the
effective radius, $r_{\rm{eff}}$. \newt{We
  compare our models to} a compilation of dwarf galaxy data in the
Local Group, including the satellites of the Milky Way and M31. The slope of the
median relation for our satellites agrees well with that of the data
for the choices $f_{\rm{MB}}=1\%$ and $3\%$. It is clear that a choice
of $5\%$ produces bright satellites that are too extended, while for
$0.5\%$ they are too compact. We therefore prefer $f_{\rm{MB}}=1\%$. A
more detailed comparison to the data at this level is problematic: the
observed sample of dwarf galaxies available at any given magnitude is
small, and the data themselves contain puzzling features such as an
apparently systematic difference in size between the bright Milky Way
and M31 satellites.

\begin{figure*}
\includegraphics[]{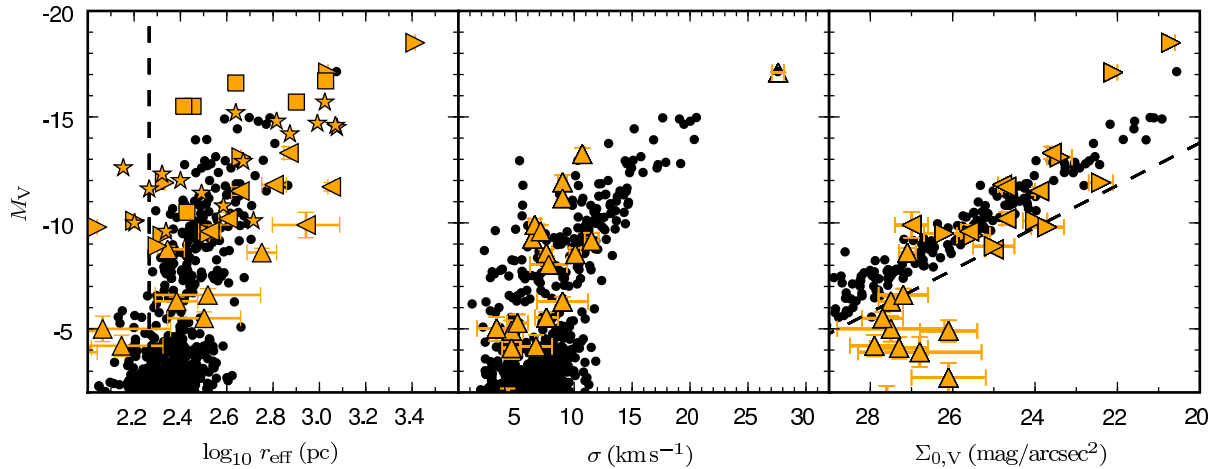}

\caption{Projected half-light radius (left), mean luminosity-weighted
  1D velocity dispersion (centre) and central surface brightness
  (right) of simulated satellite galaxies (defined by
  $r_{\rm{GC}}<280\,\rm{kpc}$) that survive in all haloes at $z=0$, as
  a function of absolute \textit{V}-band magnitude. Observational data
  for Milky Way and M31 satellites are shown as orange symbols; values
  are from Mateo (1998) and other authors as follows: bright
  satellites (triangles pointing right, Grebel, Gallagher \& Harbeck 2003); faint MW
  satellites discovered since 2005 (triangles pointing
    up, Martin, de~Jong \& Rix 2008); M31 dwarf spheroidals (triangles pointing
    left, McConnachie {et~al.} 2006; Martin {et~al.} 2009); M31 ellipticals (squares); Local
  Group `field' dwarf spheroidals and dwarf irregulars (stars). In the
  central panel we use data for Milky Way satellites only tabulated by
  Wolf {et~al.} (2009) and for the SMC, Grebel {et~al.} (2003). In the
  rightmost panel, we plot data for the Milky Way and M31
  (Grebel {et~al.} 2003; Martin {et~al.} 2008). A dashed line indicates the surface
  brightness of an object of a given magnitude with
  $r_{\rm{eff}}=2.8\epsilon$, the gravitational softening scale (see
  \mnsec{sec:aquarius}).}
\label{fig:satellite_relations}
\end{figure*}

\fig{fig:resolutiontests} also shows (as dotted lines) the same results
for our model run on the lower-resolution simulations of haloes Aq-A and
Aq-F. The particle mass in the Aq-3 series is approximately three times
greater than in Aq-2, and the force softening scale is larger by a
factor of two. We concentrate on the convergence behaviour of our
simulations for galaxies larger than the softening length, and also
where our sample provides a statistically meaningful number of galaxies
at a given magnitude; this selection corresponds closely to the regime
of the brighter dwarf spheroidal satellites of the Milky Way and M31,
$-15<M_{\rm{V}}<-5$. In this regime, \fig{fig:resolutiontests} shows
convergence of the median relations brighter than $M_{\rm{V}}=-5$ for
$f_{\rm{MB}}=3\%$ and $5\%$. The case for $f_{\rm{MB}}=1\%$ is less
clear-cut. The number of particles available for a given assignment is
set by the mass of the halo; haloes near the resolution limit (with
$\sim100$ particles) will, of course, have only $\sim1$ particle
selected in a single assignment. In addition to this poor resolution,
galaxies formed by such small-number assignments are more sensitive to
spurious two-body heating in the innermost regions of subhaloes. We
therefore expect the resulting galaxies to be dominated by few-particle
`noise' and to show poor
convergence behaviour.

We adopt $f_{\rm{MB}}=1\%$ as a reasonable match to the data (noting
also that it lies close to the power-law fit employed by
Bullock \& Johnston (2005) to map luminosities to satellite sizes). We believe
the resulting satellites to be sufficiently converged at the
resolution of our Aq-2 simulations with this choice of $f_{\rm{MB}}$
to permit a statistical study of the disrupted population represented
by the stellar halo. In support of this assertion, we offer the
following heuristic argument. The change in resolution from Aq-3 to
Aq-2 results in approximately three times more particles being
selected at fixed $f_{\rm{MB}}$; likewise, a change in $f_{\rm{MB}}$
from 1\% to 3\% selects three times more particles at fixed
resolution. Therefore, as $f_{\rm{MB}}=3\%$ has converged at the
resolution of Aq-3, it is reasonable to expect that $f_{\rm{MB}}=1\%$
selects a sufficient number of particles to ensure that satellite
sizes are not dominated by noise at the resolution of Aq-2.  We show
below that the most significant contribution to the halo comes from a
handful of well resolved objects with $M_{\rm{V}} < -10$, rather than
from the aggregation of many fainter satellites. Additionally, as
demonstrated for example by
Pe{\~{n}}arrubia,  McConnachie \& Navarro (2008a); Pe{\~{n}}arrubia,  Navarro, \& McConnachie (2008b); Pe{\~{n}}arrubia {et~al.} (2009), there is a
`knife-edge' between the onset of stellar stripping and total
disruption for stars deeply embedded within the innermost few percent
of the dark matter in a halo. We conclude that premature stripping
resulting from an over-extension of very small satellites in our model
is unlikely to alter the gross properties of our stellar haloes.

The points raised above in connection with \fig{fig:resolutiontests}
make clear that the \textit{in vivo} particle tagging approach demands
extremely high resolution, near the limits of current cosmological
N-body simulations. The choice of $f_{\rm{MB}}=1\%$ in this approach
(from an acceptable range of $1-3\%$) is not arbitrary. For example, a
choice of $f_{\rm{MB}}=10\%$ (either as a round-number estimate,

For the remainder of this paper we concentrate on the higher
resolution Aq-2 simulations. In \fig{fig:satellite_relations} we fix
$f_{\rm{MB}}$ at 1\% and compare the surviving satellites from all
six of our haloes with observational data for three properties
correlated with absolute magnitude: effective radius,
$r_{\rm{eff}}$, mean luminosity-weighted line-of-sight velocity
dispersion, $\sigma$, and central surface brightness, $\mu_{0}$
(although the latter is not independent of $r_{\rm{eff}}$). In all
cases our model satellites agree well with the trends and scatter in
the data brighter than $M_{\rm{V}} = -5$.

The force softening scale of the simulation (indicated in the first
and third panels by dashed lines) effectively imposes a maximum
density on satellite dark haloes. \newt{At} \newt{t}his radial scale
we would expect $r_{\rm{eff}}$ to become independent of magnitude
\newt{for numerical reasons}: \fig{fig:satellite_relations} shows that
the $r_{\rm{eff}}(M_{\rm{V}})$ relation becomes
steeper for galaxies fainter than
$M_{\rm{V}}\sim-9\,$, corresponding to
$r_{\rm{eff}}\sim200\,\rm{pc}$. This resolution-dependent maximum
density corresponds to a minimum surface brightness at a given
magnitude. The low-surface-brightness limit in the Milky Way data
shown in the right-hand panel of \fig{fig:satellite_relations}
corresponds to the completeness limit of current surveys
(e.g. Koposov {et~al.} 2008; Tollerud {et~al.} 2008). The lower surface brightness
satellite population predicted by our model is not, in principle,
incompatible with current data.

\begin{figure}
  \includegraphics[]{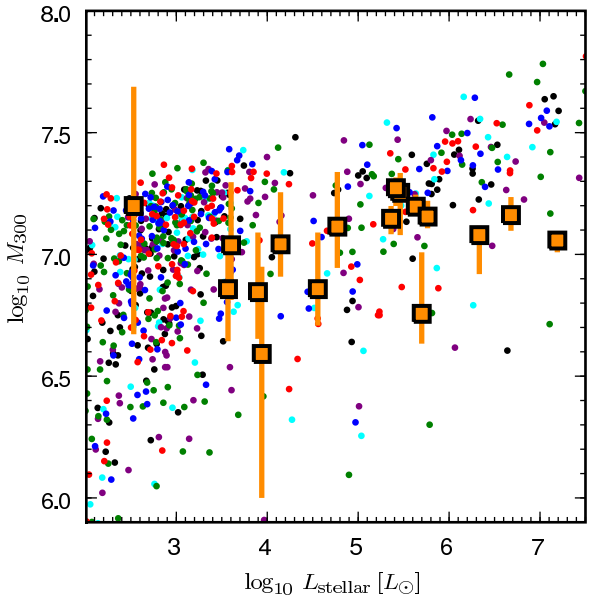}
  \caption{Mass in dark matter enclosed within 300 pc ($M_{300}$) as a
    function of luminosity (\textit{V}-band) for satellites in each of our
    simulated haloes (coloured points, colours as
    \fig{fig:galform_lf}). Maximum likelihood values of $M_{300}$ for
      Milky Way dwarf spheroidals from Strigari {et~al.} (2008) are shown
      (orange squares), with error bars indicating the range with
      likelihood greater than 60.6\% of the maximum.}
  \label{fig:m300}
\end{figure}

In \fig{fig:m300} we show the relationship between total luminosity
and the mass of dark matter enclosed within 300~pc, $M_{300}$, for our
simulated satellites in all haloes. This radial scale is well-resolved
in the level 2 Aquarius simulations (see also Font et al. 2009, in
prep.). Our galaxies show a steeper trend than the data of
Strigari {et~al.} (2008), with the strongest discrepency (0.5 dex in
$M_{300}$) for the brightest satellites. Nevertheless, both show very
little variation, having $M_{300}\sim10^{7}\,\rm{M_{\sun}}$ over five
orders of magnitude in luminosity. \newt{I}n agreement with previous
studies using semi-analytic models and lower-resolution N-body
simulations (Macci{\`{o}}, Kang \& Moore 2009; Busha {et~al.} 2009; Li {et~al.} 2009b; Koposov {et~al.} 2009), and N-body
gasdynamic simulations (Okamoto \& Frenk 2009), we find that
this characteristic scale arises naturally as a result of
astrophysical processes including gas cooling, star formation and
feedback.

\subsection{Defining the stellar halo and satellite galaxies}
\label{sec:defining_haloes}

To conclude this section, we summarise the terminology we adopt when
describing our results. Tagged dark matter particles in the
self-bound haloes and subhaloes identified by \subfind{} constitute
our `galaxies'. Our stellar haloes comprise all tagged particles
bound to the main halo in the simulation, along with those tagged
particles not in any bound group (below we impose an additional
radial criterion on our definition of the stellar halo). All
galaxies within 280~kpc of the centre of the main halo are classed
as `satellites', as in the luminosity functions shown in
\fig{fig:galform_lf}. Centres of mass of the stellar haloes and
satellites are determined from tagged particles only, using the
iterative centring process described by
Power {et~al.} (2003).

Many structural elements of a galaxy intermix within a few kiloparsecs
of its centre, and attempts to describe the innermost regions of a
stellar halo require a careful and
unambiguous definition of other components present. This is especially
important when distinguishing between those components that are
represented in our model and those that are not. Therefore, before
describing our haloes\footnote{We explicitly distinguish between the
  stellar halo and the dark halo in ambiguous cases; typically the
  former is implied throughout}, we first summarise some of these
possible sources of confusion, clarify what is and is not included in
our model, and define a range of galactocentric distances on which we
will focus our analysis of the stellar halo.

As discussed above, our model does not track with particles any
stars formed in situ in the central `Milky Way' galaxy, whether in a
rotationally supported thin disc or otherwise (this central galaxy
is, of course, included in the underlying semi-analytic model). We
therefore refer to the halo stars that \textit{are} included in our
model as \textit{accreted} and those that form in the central galaxy
(and hence are \textit{not} explicitly tracked in our model) as \textit{in
situ}. Observational definitions of the `stellar halo' typically do
not attempt to distinguish between accreted and in situ stars, only
between components separated empirically by their kinematic, spatial
and chemical properties.

The `contamination' of a purely-accreted halo by stars formed in situ
is likely to be most acute near the plane of the disc. Observations of
the Milky Way and analogous galaxies frequently distinguish a `thick
disc' component (Gilmore \& Reid 1983; Carollo {et~al.} 2009) thought to
result either from dynamical heating of the thin disc by minor mergers
(e.g. Toth \& Ostriker 1992; Quinn, Hernquist \& Fullagar 1993; Velazquez \& White 1999; Font {et~al.} 2001; Benson {et~al.} 2004; Kazantzidis {et~al.} 2008)
or from accretion debris
(Abadi {et~al.} 2003; Yoachim \& Dalcanton 2005, 2008). \movt{The presence of such a
  component in M31 is unclear: an `extended disc' is observed
  (Ibata {et~al.} 2005), which rotates rapidly, contains a young stellar
  population and is aligned with the axes of the thin disc, but
  extends to $\sim40\,\rm{kpc}$ and shows many irregular morphological
  features suggestive of a violent origin.} In principle, our model
 \newt{will follow the formation of} accreted
thick discs. However, \newt{the stars in our model only feel the
  potential of the dark halo}; the presence of a
massive baryonic disc could significantly alter this potential in the
central region and \newt{influence} the formation of \newt{an} accreted thick
disc (e.g. Velazquez \& White 1999). 

 \newt{O}ur models \newt{include that part of the galactic bulge built
   from accreted stars, but none of the many other possible processes
   of bulge formation (starbursts, bars etc.)}. However, the
 interpretation of this component, the signatures of an observational
 counterpart and the extent to which our simulation accurately
 represents its dynamics are all beyond the scope of this
 paper. \newt{Instead}, we will consider stars within \rbulge{} of the
 dark halo potential centre as `accreted bulge', and define those
 between \rbulge{} and a maximum radius of 280~kpc as the `stellar
 halo' on which we will focus our analysis. This arbitrary radial cut
 is chosen to exclude the region in which the observational separation
 of `bulge' and `halo' stars is not straightforward, and which is
 implicitly excluded from conventional observational definitions of
 the halo. It is \textit{not} intended to reflect a physical
 scale-length or dichotomy in our stellar haloes, analogous to that
 claimed for the Milky Way
 (e.g. Carollo {et~al.} 2007, 2009). Beyond \rbulge{} we
 believe that the ambiguities discussed above and the `incompleteness'
 of our models with regard to stars formed {\em in situ} should not
 substantially affect the comparison of our \textit{accreted} stars
 with observational data.

\section{Results: The Aquarius Stellar Haloes}
\label{sec:results}

\begin{figure*}
\centering \includegraphics[width=160mm]{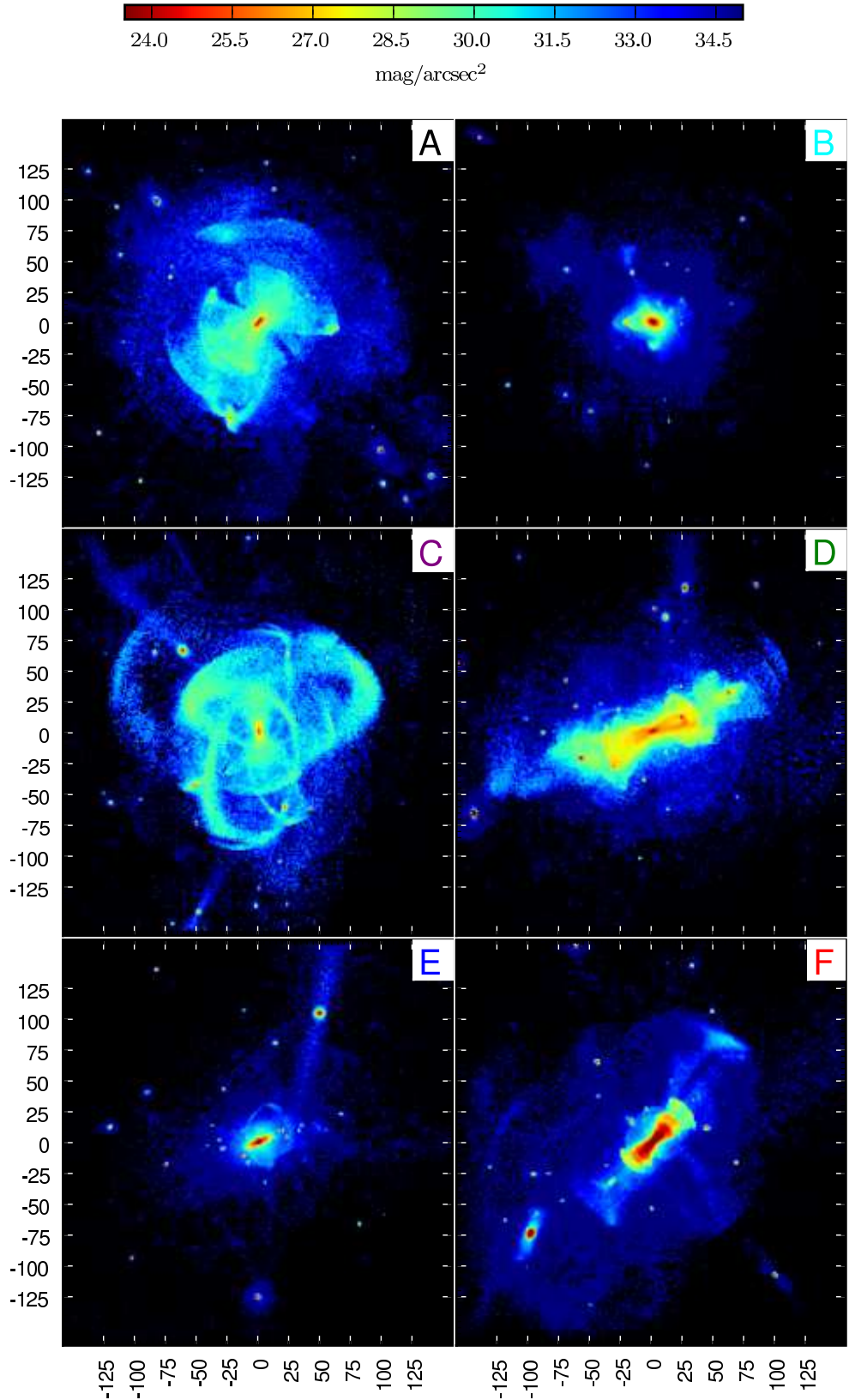}
\caption{\textit{V}-band surface
  brightness of our model haloes (and surviving satellites), to a
  limiting depth of $35\,\rm{mag/arcsec^{2}}$. The axis scales are in
  kiloparsecs. Only stars formed in satellites are present in our
  particle model; there is no contribution to these maps from a
  central galactic disc or bulge formed in situ (see
  \mnsec{sec:defining_haloes}) }
\label{fig:surfacebrightness}
\end{figure*}

\begin{figure*}
\centering
\centering \includegraphics[width=160mm]{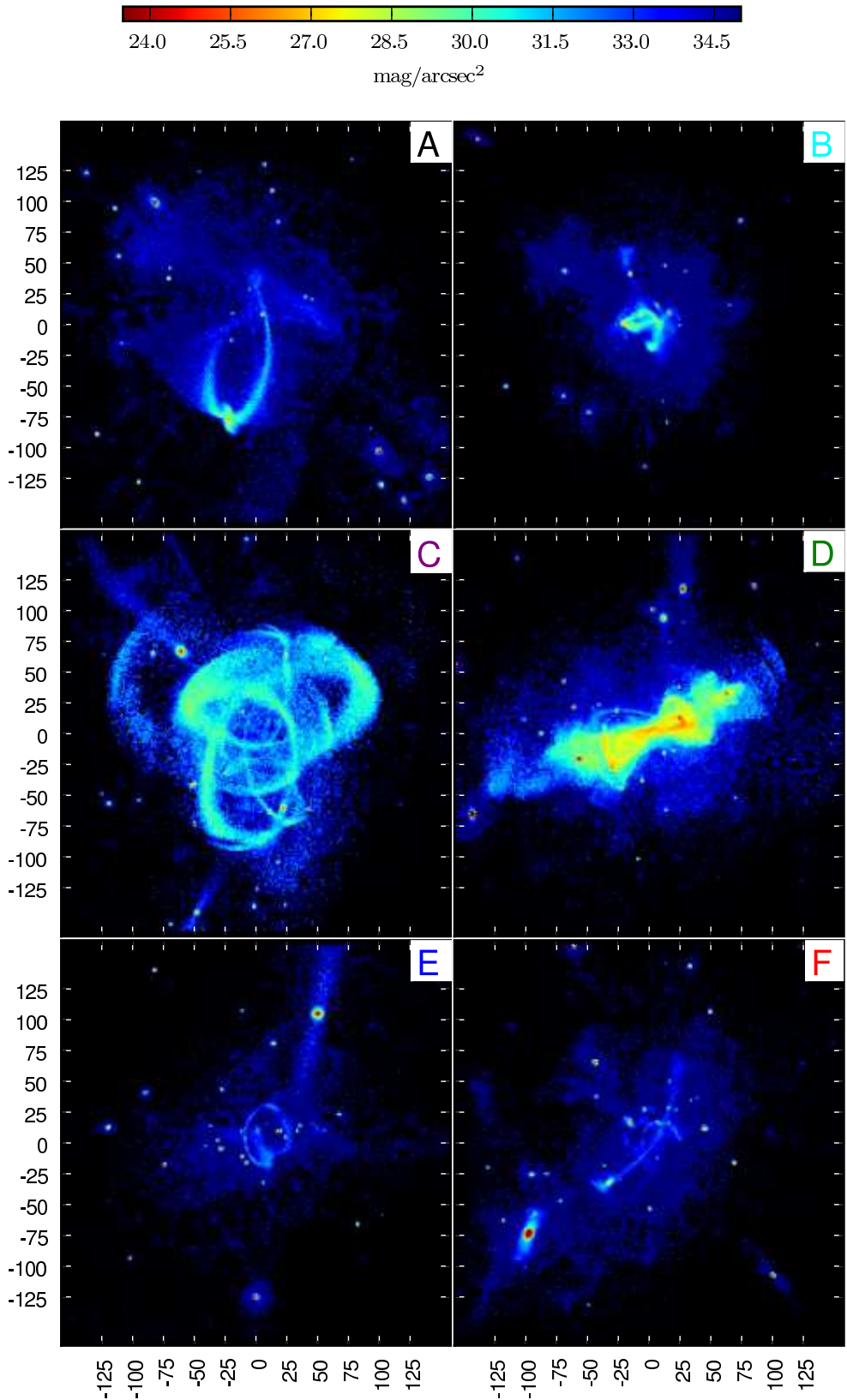}
\caption{As \fig{fig:surfacebrightness}, but here showing only those
  stars stripped from satellites that survive at $z=0$}
\label{fig:sbsurvivors}
\end{figure*}

\begin{table*}
  \begin{minipage}{160mm} \caption{For each of our simulated haloes
  we tabulate: the luminosity and mass of halo stars (in the range
  $3<r<280$~kpc); the mass of accreted bulge stars ($r<3$~kpc); the
  total stellar mass and \textit{V}-band magnitude of the central galaxy in
  \galform{}; the number of surviving satellites (brighter than
  $M_{\rm{V}}=0$); the fraction of the total stellar mass within
  280~kpc bound in surviving satellites at $z=0$, $f_{\rm{sat}}$;
  the fraction of \textit{halo} stellar mass ($r<280$~kpc)
  contributed by these surviving satellites, $f_{\rm{surv}}$; the
  number of halo progenitors, $N_{\rm{prog}}$ (see text); the
  half-light radius of the stellar halo ($r<280$~kpc); the inner and
  outer slope and break radius of a broken power-law fit to the
  three-dimensional density profile of halo stars ($3<r<280$~kpc).}
  \begin{tabular}{@{}lcccccccccccccc}
\hline
Halo      & 
$L_{V,\rm{halo}}$&
$M_{\star,\rm{halo}}$&
$M_{\star,\rm{bulge}}$&
$M_{\rm{gal}}$&
$M_{V}$&

$N_{\rm{sat}}$&

$f_{\rm{sat}}$&
$f_{\rm{surv}}$&
$N_{\rm{prog}}$&

$r_{1/2}$&
$n_{\rm{in}}$&
$n_{\rm{out}}$&
$r_{\rm{brk}}$ \\

&
$[10^{8} \rm{L_{\sun}}]$ & 

$[10^{8} \rm{M_{\sun}}]$  &     
$[10^{8} \rm{M_{\sun}}]$  &     
$[10^{10} \rm{M_{\sun}}]$ &          
& 

& 
& 
& 
&
$\rm{[kpc]}$ & & & $\rm{[kpc]}$ \\

  \hline
  A &    1.51  &    2.80  &   1.00 & 1.88 & -20.3 & 161 & 0.61 & 0.065 & 3.8 & 20  & -2.7 & -8.2 & 80.4\\
  B &    1.27  &    2.27  &   3.33 & 1.49 & -20.1 & 91  & 0.07 & 0.036 & 2.4 & 2.3 & -4.2 & -5.8 & 34.6\\
  C &    1.95  &    3.58  &   0.34 & 7.84 & -21.3 & 150 & 0.28 & 0.667 & 2.8 & 53  & -2.0 & -9.4 & 90.8\\
  D &    5.55  &    9.81  &   1.32 & 0.72 & -19.1 & 178 & 0.35 & 0.620 & 4.3 & 26  & -2.0 & -5.9 & 37.7\\
  E &    0.90  &    1.76  &  16.80 & 0.45 & -18.6 & 135 & 0.11 & 0.003 & 1.2 & 1.0 & -4.7 & -4.4 & 15.2\\
  F &   17.34  &   24.90  &   6.42 & 1.36 & -20.1 & 134 & 0.28 & 0.002 & 1.1 & 6.3 & -2.9 & -5.9 & 14.0\\
  \hline
    \end{tabular}
  \label{tbl:summary}
  \end{minipage}
\end{table*}

In this section, we present the six stellar haloes resulting from the
application of the method described above to the Aquarius
simulations. Here our aim is to characterise the assembly history of
the six haloes and their global properties. Quantities measured for
each halo are collected in \mntab{tbl:summary}. These include  a measure of the number of progenitor galaxies
contributing to the stellar halo, $N_{\rm{prog}}$. This last quantity
is not the total number of accreted satellites, but instead is defined
as $N_{\rm{prog}}=M_{\rm{halo}}^{2}/\sum_{i}{m_{\rm{{prog,i}}}^{2}}$
where $m_{\rm{{prog,i}}}$ is the stellar mass contributed by the i'th
progenitor. $N_{\rm{prog}}$ is equal to the total number of
progenitors in the case where each contributes equal mass, or to the
number of significant progenitors in the case where the remainder
provide a negligible contribution.

\subsection{Visualisation in projection}
\label{sec:projections}

A $300\times300\:\rm{kpc}$ projected surface brightness map of each
stellar halo at $z=0$ is shown in
\fig{fig:surfacebrightness}. Substantial diversity among the six
haloes is apparent. Haloes Aq-B and Aq-E are distinguished by
their strong central concentration, with few features of detectable
surface brightness beyond $\sim 20\,\rm{kpc}$. Haloes Aq-A, Aq-C,
Aq-D and Aq-F all show more extended envelopes to 75-100 kpc; each
envelope is a superposition of streams and shells that have been
phase-mixed to varying degrees.

Analogues of many morphological features observed in the halo of M31
(Ibata {et~al.} 2007; Tanaka {et~al.} 2009; McConnachie {et~al.} 2009) and other galaxies
(e.g. Mart{\'{i}}nez-Delgado {et~al.} 2008) can be found in our simulations. For
example, the lower left quadrant of Aq-A
shows arc-like features reminiscent of a complex of `parallel' streams
in the M31 halo labelled A, B, C and D by Ibata {et~al.} (2007) and
Chapman {et~al.} (2008), which have surface brightnesses of
$30-33\,\rm{mag\,arcsec^{{-}2}}$ and a range of metallicities
(Tanaka {et~al.} 2009). These streams in Aq-A can also be traced faintly in
the upper right quadrant of the image and
superficially resemble the edges of `shells'. In fact, they result
from two separate progenitor streams, each tracing multiple wraps of
decaying orbits (and hence contributing more than one `arc'
each). Seen in three dimensions, these two debris complexes (which are
among the most significant contributors to the Aq-A halo) are
elaborate and irregular structures, the true nature of which is not
readily apparent in any given projection\footnote{Three orthogonal
  projections for each halo can be found at
  \url{http://www.virgo.dur.ac.uk/aquarius}}.

The brightest and most coherent structures visible in
\fig{fig:surfacebrightness} are attributable to the most recent
accretion events. To illustrate the contribution of
recently-infalling objects (quantified in
\mnsec{sec:assembly}), we show the same projections of the haloes in
\fig{fig:sbsurvivors}, but include only those stars whose parent
satellite survives at $z=0$. In haloes Aq-C and Aq-D, stars stripped
from surviving satellites constitute $\sim60-70\%$ of the halo, while in
the other haloes their contribution is $\lesssim10\%$. Not all
the recently-infalling satellites responsible for bright halo features
survive; for example, the massive satellite that merges at $z\sim0.3$
and produces the prominent set of `shells' in Aq-F.

\begin{table}
    \caption{Axial ratios $q=c/a$ and $s=b/a$ of stellar-mass-weighted
      three-dimensional ellipsoidal fits to halo stars within a
      galactocentric radius of 10 kpc. These were determined using the
      iterative procedure described by Allgood {et~al.} (2006), which
      attempts to fit the shapes of self-consistent `isodensity'
      contours. A spherical contour of $r=10$~kpc is assumed
      initially; the shape and orientation of this contour are then
      updated on each iteration to those obtained by diagonialzing the
      inertia tensor of the mass enclosed (maintaining the length of
      the longest axis). The values thus obtained are slightly more
      prolate than those obtained from a single diagnonalization using
      all mass with a spherical contour (i.e. the first iteration of
      our approach), reflecting the extremely flattened shapes of our
      haloes at this radius. The oblate shape of Aq-E is not sensitive
      to this choice of method.}
    \begin{tabular}{@{}lcccccc}
\hline

Halo & A & B & C & D & E & F\\
\hline 
$q_{10}$ & 0.27 & 0.28 & 0.29 & 0.33 & 0.36 & 0.21\\
$s_{10}$ & 0.30 & 0.32 & 0.32 & 0.42 & 0.96 & 0.25\\

\hline
    \end{tabular}
    \label{tbl:shapes}
\end{table}

\fig{fig:surfacebrightness} shows that all our haloes are notably
flattened, particularly in the central regions where \newt{most} of
their light is concentrated. Axial ratios $q=c/a$ and $s=b/a$ of
three-dimensional ellipsoidal fits to halo stars within 10 kpc of the
halo centre are given in \mntab{tbl:shapes} (these fits include stars
within the accreted bulge region defined above). \movt{Most of
    our haloes are strongly prolate within 10~kpc. Halo Aq-E is very
    different, having a highly oblate (i.e. disc-like) shape in this
    region -- this structure of $\sim20$~kpc extent can be seen `edge
    on' in \fig{fig:surfacebrightness} \newt{and can be described as} an `accreted thick disc'
    \newt{(e.g. Abadi {et~al.} 2003; Pe{\~{n}}arrubia, McConnachie \&  Babul 2006; Read {et~al.} 2008)}.  We defer
    further analysis of this interesting object to a subsequent
    paper.}  Beyond 10--30~kpc, the stellar mass in our haloes is not
\newt{smoothly }distributed but \movt{instead} consists of a number of discrete
streams, plumes and other irregular structures. Fits to all halo stars
assuming a smoothly varying ellipsoidal distribution of mass interior
to a given radius do not accurately describe these sparse outer
regions.

Few observations of stellar halo shapes are available for comparison
with our models. M31 is the only galaxy in which a projected stellar
halo has been imaged to a depth sufficient to account for a
significant fraction of halo stars. Pritchet \& van~den Bergh (1994) measured a
projected axial ratio for the M31 halo at $\sim10$~kpc of
$\sim0.5$. Ibata {et~al.} (2005) describe a highly irregular and rotating
inner halo component or `extended disc' (to $\sim40\,\rm{kpc}$) of
$27-31\,\rm{mag/arcsec^{2}}$, aligned with the thin disc and having an
axial ratio $\sim0.6$ in projection. Zibetti \& Ferguson (2004) find a
similar axial ratio for the halo of a galaxy at $z=0.32$ observed in
the Hubble ultra-deep field. Evidence for the universality of
flattened stellar haloes is given by Zibetti, White \& Brinkmann (2004), who find a
best-fitting projected axial ratio of $\sim0.5-0.7$ for the low
surface brightness envelope of $\sim1000$ stacked edge-on late-type
galaxies in SDSS. A mildly \textit{oblate} halo with $c/a\sim0.6$ is
reported for the Milky Way, with an increase in flattening at smaller
radii ($<20$~kpc;
  e.g. Chiba \& Beers 2000; Bell {et~al.} 2008; Carollo {et~al.} 2007). Interestingly,
Morrison {et~al.} (2009) present evidence for a highly flattened halo
($c/a\sim0.2$) component in the Solar neighbourhood, which appears to
be dispersion-supported (i.e. kinematically \textit{distinct} from a
rotationally supported thick disc).

The shapes of components in our haloes selected by their kinematics,
chemistry or photometry may be very different to those obtained from
the aggregated stellar mass. A full comparison, accounting for the
variety of observational selections, projection effects and
definitions of `shape' used in the measurements cited above, is beyond
the scope of this paper. We emphasize, however, that the flattening in
our stellar haloes cannot be attributed to any `baryonic' effects such
as a thin disc potential (e.g. Chiba \& Beers 2001) or star formation in
dissipative mergers and bulk gas flows
(e.g. Bekki \& Chiba 2001). Furthermore, it is unlikely to be the result
of a (lesser) degree of flattening in the dark halo. Instead the
structure of these components is most likely to reflect the
intrinsically anisotropic distribution of satellite orbits. In certain
cases (for example, Aq-D and Aq-A), it is clear that several
contributing satellites with correlated trajectories are responsible
for reinforcing the flattening of the inner halo.

\subsection{Assembly history of the stellar halo}
\label{sec:assembly}

\begin{figure}
\includegraphics[width=84mm,clip=True, trim=0mm 0cm 0cm
  0cm]{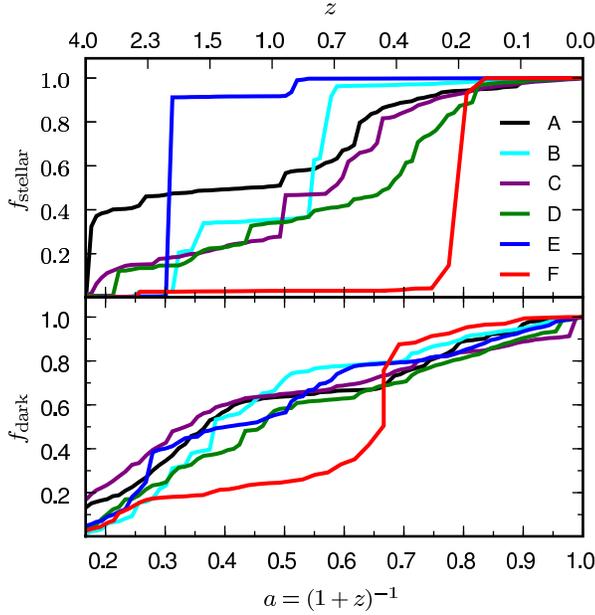}
\caption{The growth of the stellar halo (\textit{upper panel}) and the
  dark matter halo (the principal branch; \textit{lower panel}) as a
  function of expansion factor (\textit{bottom axis}) or redshift
  (\textit{top axis}). Lines show the mass fraction of each halo in
  place at a given time. Stars are counted as belonging to the stellar
  halo when the DM particle that they tag is assigned to the principal
  halo, or is not bound to any \subfind{} group.}
\label{fig:halogrowth}
\end{figure}

We now examine when and how our stellar haloes were
assembled. \fig{fig:halogrowth} shows the mass fraction of each
stellar halo (here \textit{including} the accreted bulge component
defined in \mnsec{sec:defining_haloes}) in place (i.e. unbound from
its parent galaxy) at a given redshift. We count \newt{as belonging to the stellar halo} all `star particles'
bound to the main dark halo and within 280 kpc of \newt{its} centre at $z=0$. This is compared with the growth of
the corresponding host dark haloes. \newt{O}ur sample spans a range of assembly histories for haloes
even though the halos have very similar final mass.

Not surprisingly, the growth of the dark halo is considerably
\newt{more smooth} than that of the stellar halo. The
`luminous' satellite accretion events contributing stars
\newt{are} a small subset of those that contribute to
the dark halo, which additionally accrete\newt{s} a
substantial fraction of its mass in the form of `diffuse' dark matter
\newt{(Wang et
  al. in prep.)}. As described in detail by
Pe{\~{n}}arrubia {et~al.} (2008a,b), the dark haloes of infalling
satellites must be heavily stripped before the deeply embedded stars
are removed. This gives rise to time-lags seen in \fig{fig:halogrowth}
between the major events building dark and stellar haloes.

To characterise the similarities and differences between their
histories, we subdivide our sample of six stellar haloes into two
broad categories: those that grow through the gradual accretion of
many progenitors (Aq-A, Aq-C and Aq-D) and those for which the
majority of stellar mass is contributed by only one or two major
events (Aq-B, Aq-E and Aq-F). We refer to this latter case as
`few-progenitor' growth. The measure of the number of
`most-significant' progenitors given in \mntab{tbl:summary},
$N_{\rm{prog}}$, also ranks the haloes by the `smoothness' of their
accretion history, reflecting the intrinsically stochastic nature of
their assembly.

\begin{figure}
\includegraphics[width=84mm,clip=True, trim=0cm 0cm 0cm 0cm]{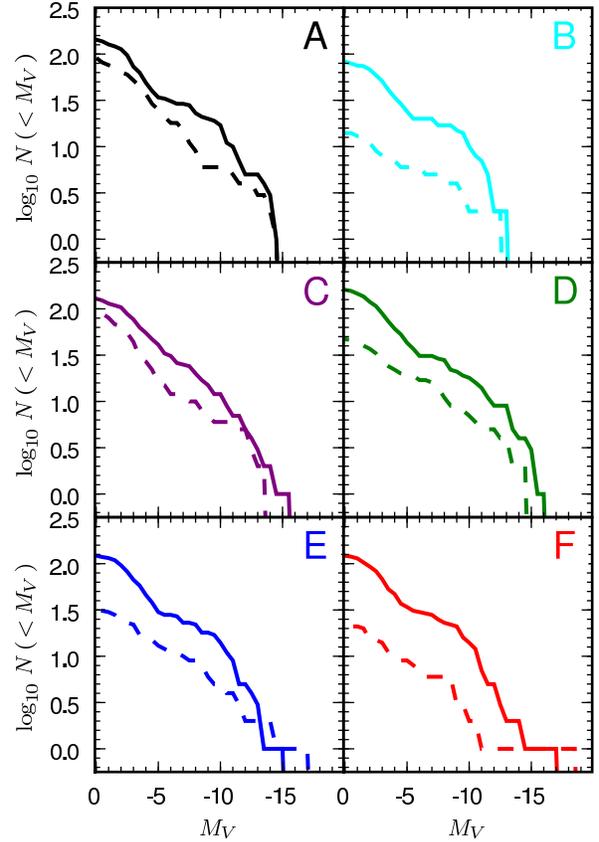}
\caption{Luminosity functions of surviving satellites (solid) in each
  of our six haloes, compared with those of totally disrupted halo
  progenitors (dashed). These are constructed using only stars formed
  in each satellite before the time of infall (the halo-subhalo
  transition). The luminosity of each population is that after
  evolution to $z=0$.}
\label{fig:victim_vs_survivor_lf}
\end{figure}

\fig{fig:victim_vs_survivor_lf} compares the luminosity
functions (LFs) of surviving satellites with \newt{that of those} totally disrupted \newt{to form the
  stellar halo}, measuring luminosity at the time of infall in
both cases. In general, there are fewer disrupted satellites than
survivors over almost all luminosities, although the numbers and
luminosities of the very brightest contributors and survivors are
comparable in each halo. The deficit in the number of disrupted
satellites relative to survivors is most pronounced in the
few-progenitor haloes Aq-B and Aq-F.
 
\begin{figure} 
\includegraphics[width=84mm,clip=True,trim=0 0 0 0]{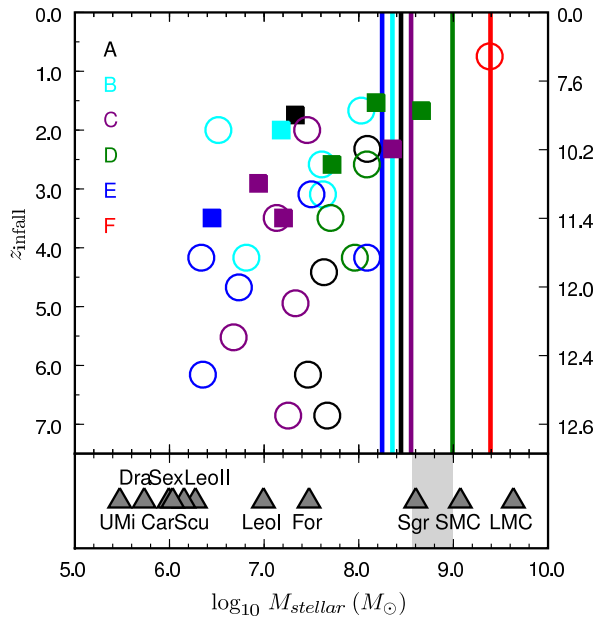}
\caption{\textit{Main panel}: for satellites that have been stripped
  to form the stellar haloes, symbols show the redshift of infall and
  total mass contributed to the stellar halo at $z=0$ (in the range $3
  < r < 280\,\rm{kpc}$). Vertical lines indicate the total mass of
  each stellar halo in this radial range. The right-hand $y$-axis is
  labelled by lookback time in gigayears. We plot only those
  satellites whose individual contributions, accumulated in rank order
  from the most significant contributor, account for 95\% of the total
  stellar halo mass. Satellites totally disrupted by $z=0$ are plotted
  as open circles, surviving satellites as filled squares (in almost
  all cases the contributions of these survivors are close to their
  total stellar masses; see text). \textit{ Lower panel:} symbols
  indicate the approximate masses of bright MW satellites, assuming a
  stellar mass-to-light ratio of 2; the Sgr present-day mass estimate
  is that given by Law, Johnston \& Majewski (2005). The shaded region indicates an
  approximate range for the MW halo mass in our halo regime (see
    e.g. Bell {et~al.} 2008).}

\label{fig:contributions}
\end{figure}

\fig{fig:contributions} summarises \movt{the
  individual accretion events that contribute to the assembly of the
  stellar halo}, plotting the stellar mass of the most significant
progenitor satellites against their redshift of infall (the
time at which their host halo first \newt{becomes} a
subhalo of the main FOF group). Here we class as significant those
satellites which together contribute 95\% of the total halo stellar
mass (this total is shown as a vertical line for each halo) when
accumulated in rank order of their contribution. By this measure there
are (5,6,8,6,6,1) significant progenitors for haloes (A,B,C,D,E,F). We
also compare the masses of the brightest Milky Way satellites to the
significant contributors in our stellar haloes. Typically the most
significant contributors have masses comparable to the most massive
surviving dwarf spheroidals, Fornax and Sagittarius.

\begin{figure}
\includegraphics[width=84mm]{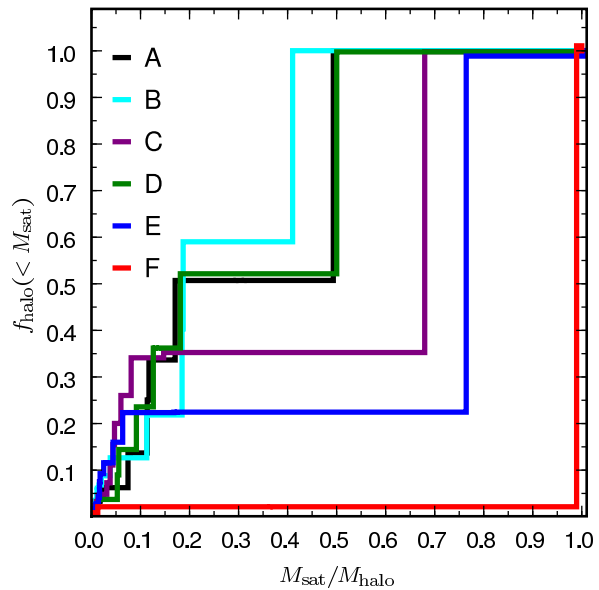}
\caption{Cumulative mass fraction of each stellar halo originating in
  satellites of stellar mass less than $M_{\rm{sat}}$. Satellite
  masses are normalised to the total stellar halo mass
  $M_{\rm{halo}}$ in each case, as defined in
  \mnsec{sec:defining_haloes}.}
\label{fig:carlos_contribs}
\end{figure}

\begin{figure}
\includegraphics[width=84mm]{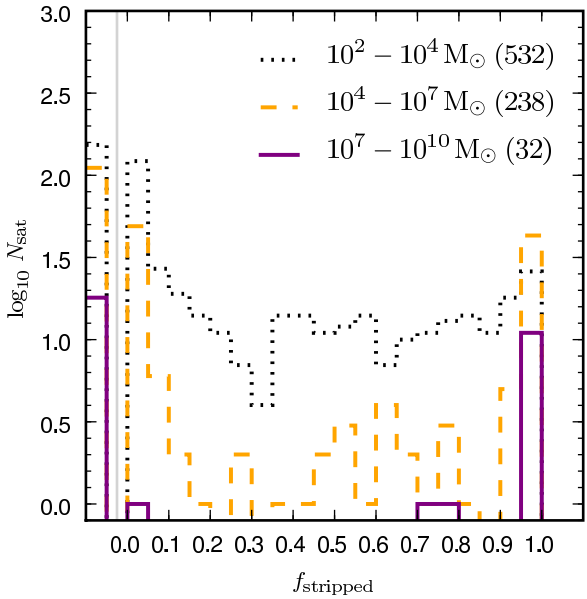}
\caption{Number of surviving satellites (aggregated over all six
  haloes) which have lost a fraction, $f_{\rm{stripped}}$, of the
  stellar mass through tidal stripping. Satellites are divided into
  three mass bins: massive (purple), intermediate (dashed orange)
  and low-mass (dotted black) as quantified in the legend. The
  leftmost bin (demarcated by a vertical line) shows the number of
  satellites that have not suffered any stellar mass loss.}
\label{fig:stripping}
\end{figure}

With the exception of Aq-F, all the most significant contributors to
our stellar haloes were accreted more than 8 Gyr ago. We highlight (as
filled squares) those contributors whose cores survive as self-bound
objects at $z=0$. We find that surviving satellites accreted before
$z=1$ are the dominant contributors to the many-progenitor haloes Aq-C
and Aq-D. The extreme case of Aq-F is atypical: more than 95\% of the
halo was contributed by the late merger of an object of stellar mass
greater than the SMC infalling at $z\sim0.7$, which does not
survive. By contrast, the two least massive haloes Aq-B and Aq-E are
built by many less massive accretions at higher redshift, with
surviving satellites making only a minor contribution ($<10\%$). Halo
Aq-A represents an intermediate case, in which stars stripped from a
relatively late-infalling survivor add significantly ($\sim10\%$) to the mass of
a halo predominantly assembled at high redshift. The relative
contributions to the halo of all accretion events are illustrated in
\fig{fig:carlos_contribs}. Each line in this figure indicates the
fraction of the total halo stellar mass that was contributed by
satellites donating less than a given fraction of this total
\textit{individually}. An interesting feature illustrated by this
figure concerns Aq-B, one of our few-progenitor haloes (shown as light
blue in all figures). Although \fig{fig:halogrowth} shows that the
assembly of this halo proceeds over time by a series of concentrated
`jumps' in mass, its final composition is even less biased to the most
significant progenitor than any of the many-progenitor haloes.

In general, surviving contributors to the halo retain less than 5\% of
the total stellar mass that formed in them. A small number of surviving contributors
retain a significant fraction of their mass, for example, the
surviving contributor to Aq-A, which retains 25\%. In
\fig{fig:stripping}, we show histograms of the number of all surviving
satellites (combining all six haloes) that have been stripped of a
given fraction of their mass. Most satellites are either largely
unaffected or almost totally stripped, indicating that the time spent
in an intermediate disrupting state is relatively short.

In \mntab{tbl:summary}, we give the fraction of mass in the stellar
halo that has been stripped from surviving satellites,
$f_{\rm{surv}}$. As previously stated, this contribution is dominant
in haloes Aq-C (67\%) and Aq-D (62\%), significant in Aq-A (7\%) and
Aq-B (4\%) and negligible in Aq-E and Aq-F. Sales {et~al.} (2007b)
find that only $\sim6\%$ of stars in the eight haloes formed in the
SPH simulations of
Abadi {et~al.} (2006) are associated with a surviving satellite. The lack of
surviving satellites may be attributable to the limited resolution of
those simulations; clearly, the number of `survivors' is sensitive to
the lowest mass at which remnant cores can be resolved. However,
Bullock \& Johnston (2005), and the companion study of Font {et~al.} (2006), also
conclude that the contribution of surviving satellites is small
($<10\%$ in all of their 11 haloes and typically $<1\%$). As the
resolution of their simulations is comparable to ours, the
predominance of surviving contributors in two of our haloes is
significant.

Bullock \& Johnston find that their haloes are built from a similar
(small) number of massive objects to ours (e.g. figure 10
of Bullock \& Johnston 2005) with comparable accretion times ($>8$ Gyr), suggesting
that there are no fundamental differences in the infall times and masses
of accreted satellites. Notably, Font {et~al.} (2006) observe that no
satellites accreted $>9$~Gyr ago survive in their subsample of four of
the Bullock \& Johnston haloes, whereas we find that some satellites
infalling even at redshifts $z>2$ may survive (see also
\fig{fig:zinfall_gradient} below). The discrepancy appears to stem from
the greater resilience of satellites accreted at $z>1$ in our models,
including some which contribute significantly to the stellar haloes. In
other words, our model does not predict any \newt{more} late-infalling
contributors \newt{than the models of Bullock \& Johnston}. The more
rapid disruption of massive subhaloes in the Bullock \& Johnston
models may be attributable to one or both of the analytic prescriptions
employed by those authors to model the growth of the dark matter halo
and dynamical friction in the absence of a live halo. It is also
possible that the relation between halo mass and concentration assumed
in the Bullock \& Johnston model results in satellites that are less
concentrated than subhaloes in the Aquarius simulations.

Current observational estimates (e.g. Bell {et~al.} 2008) imply that the
stellar halo of the Milky Way is intermediate in mass between our
haloes Aq-C and Aq-D; if its accretion history is, in fact, 
qualitatively similar to these many-progenitor haloes,
\fig{fig:contributions} implies that it is likely to have accreted its
four or five most significant contributors around $z\sim1-3$ in the
form of objects with masses similar to the Fornax or Leo I dwarf
spheroidals. Between one and three of the most recently accreted, and
hence most massive contributors, are expected to retain a surviving
core, and to have a stellar mass comparable to Sagittarius
($M_{\rm{sgr}} \sim 5 \times 10^{8}\,\rm{M_{\sun}}$ or $\sim 50\%$ of
the total\footnote{Both the Sagittarius and Milky Way halo stellar
  mass estimates are highly uncertain; it is unclear what contribution
  is made by the Sgr debris to estimates of the halo mass, although
  both the stream and the Virgo overdensity were masked out in the
  analysis of Bell {et~al.} (2008) for which a value of $\sim3\times
  10^{8}\,\rm{M_{\sun}}$ in the range $3<r<40\,\rm{kpc}$ was obtained from a
  broken power-law fit to the remaining `smooth' halo.} halo mass,
infalling at a lookback time of $\sim5\,\rm{Gyr}$;
Law {et~al.} 2005). It is also possible that the Canis Major
overdensity (with a core luminosity comparable to that of
Sagittarius; Martin {et~al.} 2004) associated with the low-latitude Monoceros
stream (Newberg {et~al.} 2002; Yanny {et~al.} 2003; Ibata {et~al.} 2003) should be
included in the census of `surviving contributors' (although
this association is by no means certain;
e.g. Mateu {et~al.} 2009). Therefore, the picture so far established for the
Milky Way appears to be in qualitative agreement with the presence
of surviving cores from massive stellar halo contributors in our
simulations.

\subsection{Bulk halo properties and observables}
\label{sec:global_haloes}

\subsubsection{Distribution of mass}
\label{sec:distribution}

\begin{figure}
\includegraphics[width=84mm,clip=True, trim=0mm 0cm 0cm 0cm]{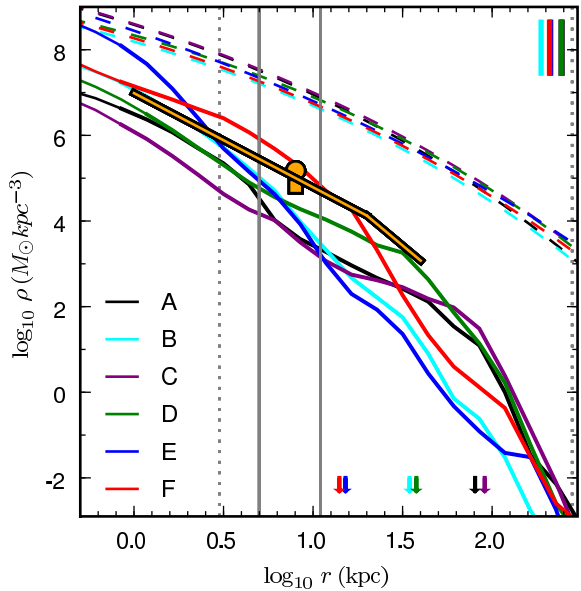}
\caption{Spherically averaged density profiles for our six stellar
  haloes (shown as thin lines below the $\kappa=7$ radius of
  Navarro {et~al.} 2008, at which the circular velocity
  of the dark matter halo has converged to an accuracy of 1\%). Arrows
  mark the break radii of broken power-law fits to each
  profile. Dashed lines show Einasto profile fits to the corresponding
  dark matter haloes (Navarro {et~al.} 2008). Grey vertical
  lines demarcate our outer halo region (dotted) and the Solar
  neighbourhood (solid); coloured vertical bars indicate $r_{200}$ for
  the dark haloes. For reference we overplot representative data for
  the Milky Way (orange): estimates of the halo density in the Solar
  neighbourhood (symbols) from Gould, Flynn \& Bahcall (1998, square) and
  Fuchs \& Jahrei{\ss{}} (1998, circle), and the best-fitting broken
  power-law of Bell et al. (excluding the Sagittarius stream and Virgo
  overdensity).}
\label{fig:densityprofiles}
\end{figure}

In \fig{fig:densityprofiles} we show the spherically averaged
density profiles of halo stars (excluding material bound in
surviving satellites, but making no distinction between streams,
tidal tails or other overdensities, and a `smooth' component). The
notable degree of substructure in these profiles contrasts with the
smooth dark matter haloes, which are well-fit by the Einasto
profiles shown in
\fig{fig:densityprofiles}. As discussed further below, this stellar
substructure is due to the contribution of localised, spatially
coherent subcomponents within the haloes, which are well resolved in
our particle representation.

The shapes of the density profiles are broadly similar, showing a
strong central concentration and an outer decline considerably steeper
than that of the dark matter. We overplot in \fig{fig:densityprofiles}
an approximation of the Milky Way halo profile (Bell {et~al.} 2008) and
normalization (Fuchs \& Jahrei{\ss{}} 1998; Gould {et~al.} 1998). The gross structure of
our three many-progenitor haloes Aq-A, Aq-C and Aq-D can be fit with
broken power-law profiles having indices similar to the Milky Way
($n\sim-3$) interior to the break. Bell {et~al.} (2008) note that their
best-fitting observational profiles do not fully represent the complex
structure of the halo, even though they mask out known overdensities
(our fits include all halo substructure). Our fits decline somewhat
more steeply than the Bell {et~al.} data beyond their break
radii. We suggest that the Milky Way fit may represent variation at
the level of the fluctuations seen in our profiles, and that an even
steeper decline may be observed with a representative and well-sampled
tracer population to $>100$ kpc (For example, Ivezi{\'c} {et~al.} 2000, find a sharp
  decline in counts of RR Lyr stars beyond $\sim60$~kpc). In
contrast with the many-progenitor haloes, two of our few-progenitor
haloes (Aq-B and Aq-E) have consistently steeper profiles and show no
obvious break. Their densities in the Solar shell are \newt{none the
  less} comparable to the many-progenitor haloes. Aq-F is dominated by
a single progenitor, the debris of which retains a high degree of
unmixed structure at $z=0$ (see also \fig{fig:deposition_profile}).

\begin{figure}
\includegraphics[width=84mm]{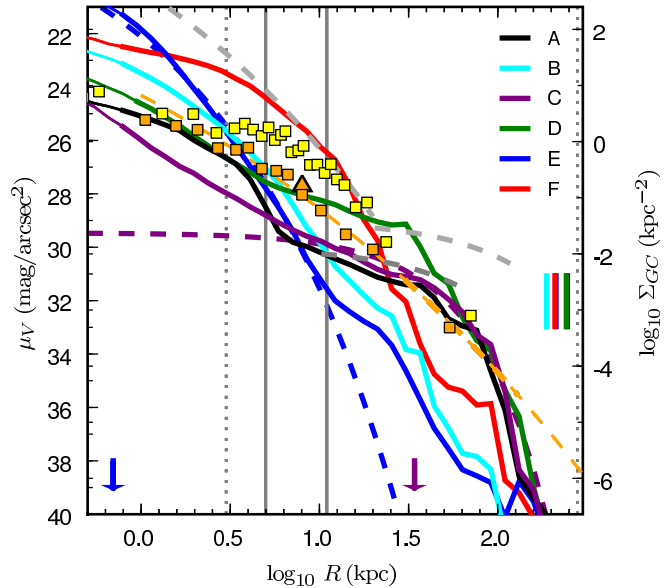}
\caption{Radially averaged surface brightness profiles. Dashed lines
  show illustrative Sersic fits to haloes Aq-E and Aq-C (see text),
  with arrows indicating the corresponding scale radii. We show
  sections of equivalent profiles for the haloes of M31 (including the
  inner $r^{1/4}$ `spheroid') and M33 (beyond 10~kpc) as dashed grey
  lines (Ibata {et~al.} 2007). We overplot the surface number density
  (right-hand axis) of globular clusters in M31 (yellow squares) and
  the Milky Way (orange squares), with 40 and 10 clusters per bin,
  respectively. These profiles have been arbitrarily normalized to
  correspond to an estimate of the surface brightness of halo stars in
  the Solar neighbourhood from Morrison (1993), shown by a
  orange triangle. Vertical lines are as in \fig{fig:densityprofiles}}
\label{fig:radialprofiles}
\end{figure}

\begin{figure}
\includegraphics[]{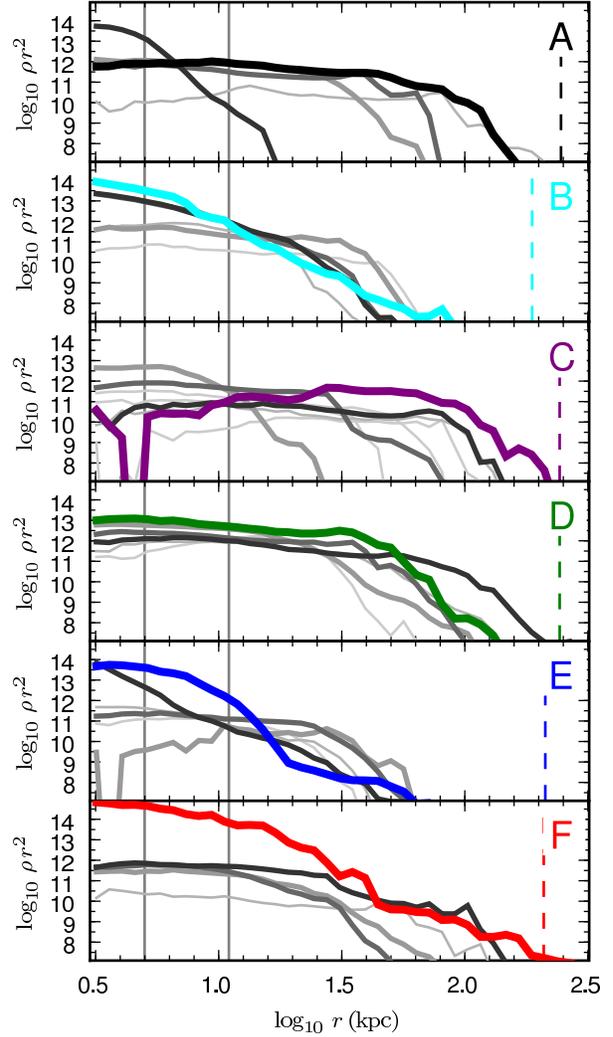}
\caption{Individual density profiles (multiplied by $r^2$) for stars
  contributed by each of the most significant progenitors of the
  halo (defined in \mnsec{sec:defining_haloes}). Line types indicate
  the rank order of a progenitor contribution: the bold coloured
  line in each panel indicates the most significant contributor,
  while lesser contributions are shown by increasingly lighter and
  thinner lines. Vertical solid and dashed lines indicate the Solar
  shell and virial radius respectively, as
  \fig{fig:densityprofiles}. Individual stellar halo components
  contribute over a wide radial range, and different components
  `dominate' at particular radii. This figure can be used to
  interpret the radial trends shown in other figures.}
\label{fig:deposition_profile}
\end{figure}

We show projected surface brightness profiles in
\fig{fig:radialprofiles}. As with their three-dimensional
counterparts, two characteristic shapes distinguish the many- and
few-progenitor haloes. The few-progenitor haloes are centrally
concentrated and well fit in their innermost $\sim10\,\rm{kpc}$ by
Sersic profiles with $1.5<n<2.2$. Beyond $10\,\rm{kpc}$, extended
profiles with a more gradual rollover (described by Sersic profiles
with $n\sim1$ and $25<r_{\rm{eff}}<35$ kpc) are a better fit to the
many-progenitor haloes. In their centres, however, the many-progenitor
haloes display a steep central inflection in surface brightness. As a
consequence of these complex profiles, Sersic fits over the entire
halo region (which we defined to begin at 3~kpc) are not fully
representative in either case. To illustrate this broad dichotomy in
\fig{fig:radialprofiles}, Sersic fits to a smoothly growing halo
(Aq-C) \textit{beyond} 10~kpc and a few-progenitor halo (Aq-E)
\textit{interior to} 10~kpc are shown.  Abadi {et~al.} (2006) found the
average of their simulated stellar haloes to be well-fit by a Sersic
profile ($n=6.3,\,r_{eff}=7.7\,$~kpc) in the radial range
$30<r<130$~kpc, \newt{which we} show as an orange dashed line
in \fig{fig:radialprofiles}. This profile is close to the `mean'
profile of our halos A, C and D interior to $30$~kpc (neglecting the
significant fluctuations and inflections within each individual halo
in \fig{fig:radialprofiles}), but does not capture the sharp decline
of our haloes at radii beyond 150~kpc. \fig{fig:radialprofiles} also
shows (as dashed grey lines) the fits of Ibata {et~al.} (2007) to the haloes
of M31 (comprising an $r^{1/4}$ spheroid and shallow powerlaw tail at
large radii) and M33 (powerlaw tail only).

There is evidence for multiple kinematic and chemical subdivisions
within the Galactic globular cluster population (e.g. Searle \& Zinn 1978; Frenk \& White 1980; Zinn 1993; Mackey \& Gilmore 2004, and
  refs. therein). This has led
to suggestions that at least some of these cluster subsets may have
originated in accreted satellites
(Bellazzini, Ferraro \&  Ibata 2003; Mackey \& Gilmore 2004; Forbes, Strader, \& Brodie 2004). Support for this conclusion
includes the presence of five globular clusters in the Fornax dwarf
spheroidal (Hodge 1961) and the association of several Galactic
clusters with the Sagittarius nucleus and debris
(e.g. Layden \& Sarajedini 2000; Newberg {et~al.} 2003; Bellazzini {et~al.} 2003). Similarities with the
`structural' properties of stellar populations in the halo have
motivated a longstanding interpretation of globular clusters as halo
(i.e. accretion debris) tracers (e.g. Lynden-Bell \& Lynden-Bell 1995). We
therefore plot in \fig{fig:radialprofiles} the surface density profile
of globular clusters in the Milky Way (Harris 1996) and M31
(confirmed GCs in the Revised Bologna Catalogue -- RBC v3.5,
  March 2008 Galleti {et~al.} 2004, 2006, 2007; Kim {et~al.} 2007; Huxor {et~al.} 2008). The Milky Way data have
been projected along an arbitrary axis, and the normalization has been
chosen to match the surface density of Milky Way clusters to an
estimate of the surface brightness of halo stars in the Solar
neighbourhood ($\mu_{V}=27.7\,\rm{mag/arcsec^2}$;
Morrison 1993). We caution that the RBC incorporates data from
ongoing surveys as it becomes available: the M31 GC profile shown here
is therefore substantially incomplete, particularly with regard to the
sky area covered beyond $\sim20$--30~kpc.

Abadi {et~al.} (2006) showed that their average stellar halo Sersic fit also
approximates the distribution of globular clusters in the Milky Way
and M31. As stated above, the inner regions of our haloes Aq-A, Aq-C
and Aq-D are in broad agreement with the Abadi {et~al.} halo
profile, and hence show some similarities with the observed globular
cluster profiles also. Both the halo and cluster samples show strong
variations from halo to halo, however, and the comparison of these
small samples is inconclusive. A close correspondence between accreted
halo stars and globular clusters would be expected only if the
majority of clusters are accreted, if accreted satellites contribute a
number of clusters proportionate to their stellar mass, and if all
stripped clusters have an equal probability of surviving to
$z=0$. None of these assumptions is realistic, and further work is
required better to constrain the relationship between globular
clusters and stellar haloes.

The multicomponent nature of our haloes, which gives rise to the local
structure in their overall profiles, is examined in more detail in
\fig{fig:deposition_profile}. Here the density profiles of the major
contributors shown in \fig{fig:contributions} are plotted
individually (progenitors contributing $<5\%$ of the halo have been
added to the panel for Aq-F). It is clear from these profiles that
material from a given progenitor can be deposited over a wide range
of radii. The few-progenitor haloes show strong gradients in $\rho
r^2$ while more uniform distributions of this quantity are seen in
their sub-dominant contributors and in most contributors to the
many-progenitor haloes.

\begin{figure}
\includegraphics[]{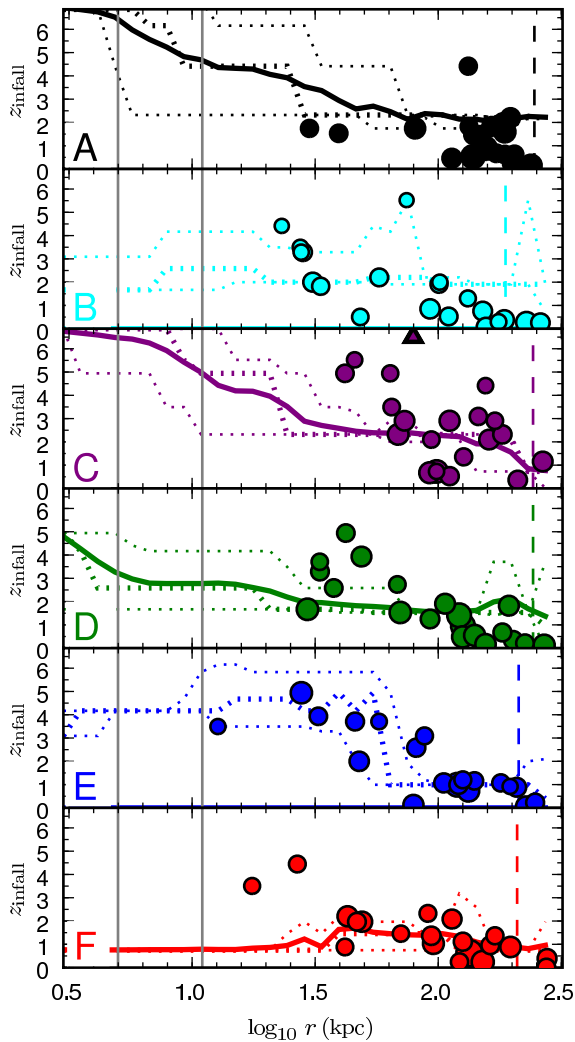}
\caption{Lines show, for halo stars at a given radius at $z=0$, the
  mean (solid), median and 10/90th percentile (dotted) redshift at
  which their parent galaxy was accreted on to the main halo
  (\textit{not} the time at which the stars themselves were
  stripped). Filled circles show the redshift at which surviving satellites
  were accreted; triangles indicate satellites accreted before
  $z=7$. Within the solar shell, the stellar halo is typically old in
  this `dynamical' sense, whereas beyond 100 kpc its young `dynamical'
  age is comparable to that of the surviving satellite population. In
  many cases the innermost satellites represent a relic population
  that is `older' than the stellar halo at comparable radii.}
\label{fig:zinfall_gradient}
\end{figure}

Finally, we show in \fig{fig:zinfall_gradient} the time at which the
satellite progenitors of halo stars at a given radius were accreted
(this infall time is distinct from the time at which the stars
themselves were stripped, which may be considerably later). An
analogous infall time can be defined for the surviving satellites,
which are shown as points in
\fig{fig:zinfall_gradient}. \newt{W}e would expect
little information to be encoded in an instantaneous sample of the
radii of surviving satellites, \newt{but} their infall times can none
the less be usefully compared with those of halo stars.

A gradient to earlier infall times with decreasing radius is apparent
in both the satellites and the many-progenitor haloes. In the case
of the haloes, this reflects the fact that relatively larger apocentres
are associated with later-infalling satellites, which enable them to
deposit material over a greater radial range. Assembly in this manner
is arguably not adequately characterised as `inside out' formation;
late infalling material is added at all radii but has a greater
maximum extent than earlier-infalling material. The result is that
earlier-infalling material comes to \textit{dominate} towards the
centre. For the few-progenitor haloes the profile of infall time is
essentially flat (or shows sharp transitions between populations),
more closely reflecting the contributions of individual progenitors.

Further to our discussion of satellite survival in our haloes in
\mnsec{sec:assembly}, it is interesting that amongst the surviving
satellites, we observe several accreted at $z>1$. For example, in
the case of Aq-E, six surviving satellites are accreted at
$z\sim3.5$; at the present day this group is found in association
with a concentration of halo stars from a stellar halo progenitor
also infalling at this time. The majority of survivors in each halo
are accreted recently, however, and typically more recently than the
stellar halo progenitors. The opposite is true for the
earliest-accreted survivors, which are accreted earlier than the
halo at the notably small radii at which they are now found. In
general, at any given instant the majority of satellites are more
likely to be located nearer to the apocentre of their orbit than the
pericentre; furthermore, the orbits of the most massive satellites
are likely to have been more circular than their disrupted siblings
and dynamical friction may act to reinforce such a trend. Therefore,
the locations of early-infalling survivors are likely to be fairly
represented by their radius in \fig{fig:zinfall_gradient}. Dynamical
friction acts to contract but also to circularize orbits. Plausibly
these survivors are those that have sunk slowly as the result of
their initially low orbital eccentricities.

\subsubsection{Stellar populations}
\label{sec:stellarpops}

In this section, we show how the multicomponent nature of our stellar
haloes is reflected in their metallicity profiles, and contrast the
stellar populations of surviving satellites with those of halo
progenitors. We caution that a full comparison of the relationship
between the stellar halo and surviving satellites will require more
sophisticated modelling of the chemical enrichment process than is
included in our fiducial model, which adopts the instantaneous
recycling approximation and does not follow individual elemental
abundances. We will address this detailed chemical modelling and
\newt{related observational comparisons} in a subsequent
paper (De Lucia et al. in prep.). The model we adopt here tracks only
total metallicity, defined as the total mass fraction of all metals
relative to the Solar value, $Z/\rm{Z_{\sun}}$ (the absolute value of
which cannot be compared directly with measurements of \feh{}). This
model can nevertheless address the \textit{relative} enrichment
levels of different populations.

\begin{figure}
\includegraphics[width=84mm]{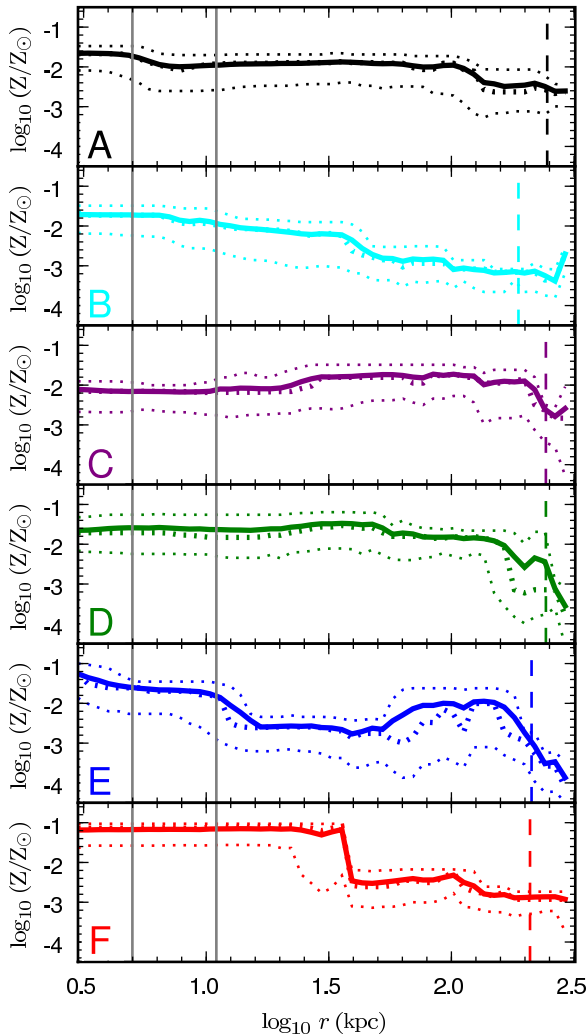}
\caption{Radial profiles of luminosity-weighted metallicity (ratio of
  total metal mass fraction to the Solar value) for spherical shells
  in our six haloes, showing the mean (solid) and median
  (thick dotted) profiles, bracketed by the 10th and 90th
  percentiles (dotted).}
\label{fig:metal_gradient}
\end{figure}

\fig{fig:metal_gradient} shows the spherically averaged metallicity
gradient in each halo. Our many-progenitor haloes are characterised by
a metallicity distribution of width $\sim1$ dex and approximately
constant mean value, fluctuating by less than $\pm0.5$ dex over a
range of 100~kpc. This is comparable to observations of the M31 halo,
which show no significant gradient (metallicities varying by $\pm0.14$
dex) in the range 30--60~kpc (Richardson {et~al.} 2009). Localised structure
is most apparent in the few-progenitor haloes: Aq-F shows a clear
separation into two components, while Aq-B and Aq-E exhibit global
trends of outwardly declining metallicity gradients. In all cases the
mean metallicity within the Solar radius is relatively high. These
features can be explained by examining the relative weighting of
contributions from individual progenitors at a given radius, as shown
in the density profiles of \fig{fig:deposition_profile}, bearing in
mind the mass-metallicity relation for satellites that arises in our
model. Where massive progenitors make a significant
luminosity-weighted contribution, the haloes are seen to be
metal-rich. Overall, metallicity gradients are shallower in those
haloes where many significant progenitors make a comparable
contribution, smoothing the distribution over the extent of the
halo. Conversely, metallicity gradients are steeper where only one or
two disproportionately massive satellites make contributions to the
halo (as indicated by the luminosity functions of
\fig{fig:victim_vs_survivor_lf}). Sharp contrasts are created between
the radii over which this metal-rich material is deposited (massive
satellites suffer stronger dynamical friction and sink more rapidly,
favouring their concentration at the centres of haloes) and a
background of metal-poor material from less massive halo
progenitors. This effect is clearly illustrated by the sharp
transition in Aq-F and at two locations (centrally and at
$\sim100\,\rm{kpc}$) in Aq-E.

It follows that the process by which our smooth haloes are
assembled, which gives rise to the steep gradients of progenitor
infall time with redshift shown in \fig{fig:zinfall_gradient}, also
acts to \textit{erase} metallicity gradients. As a result,
measurements of (for example) \feh{} alone do not constrain the local
infall time; a metal-poor halo need not be `old' in the sense of early
assembly. A particularly notable example of this is Aq-E, where the
centrally dominant metal-rich material was assembled into the halo
considerably \textit{earlier} ($z\sim3$) than the diffuse outer
envelope of relatively metal-poor material ($z\sim1$). This is a
manifestation of a mass-metallicity relation in satellites: at fixed
luminosity, an earlier infall time is `compensated' for by more rapid
star formation, resulting in a comparable degree of overall enrichment
as that for a satellite with similar luminosity infalling at lower
redshift. Abundance ratios such as \afe{} indicate the time taken by a
given stellar population to reach its observed level of enrichment,
and so distinguish between rapidly forming massive populations,
truncated by early accretion to the halo, and populations reaching
similar mass and metallicity through gradual star formation
(e.g. Shetrone, C{\^{o}}t{\'{e}} \&  Sargent 2001; Tolstoy {et~al.} 2003; Venn {et~al.} 2004; Robertson {et~al.} 2005).

\begin{figure}
\includegraphics[clip=True, trim=0mm 0cm 0cm 0cm]{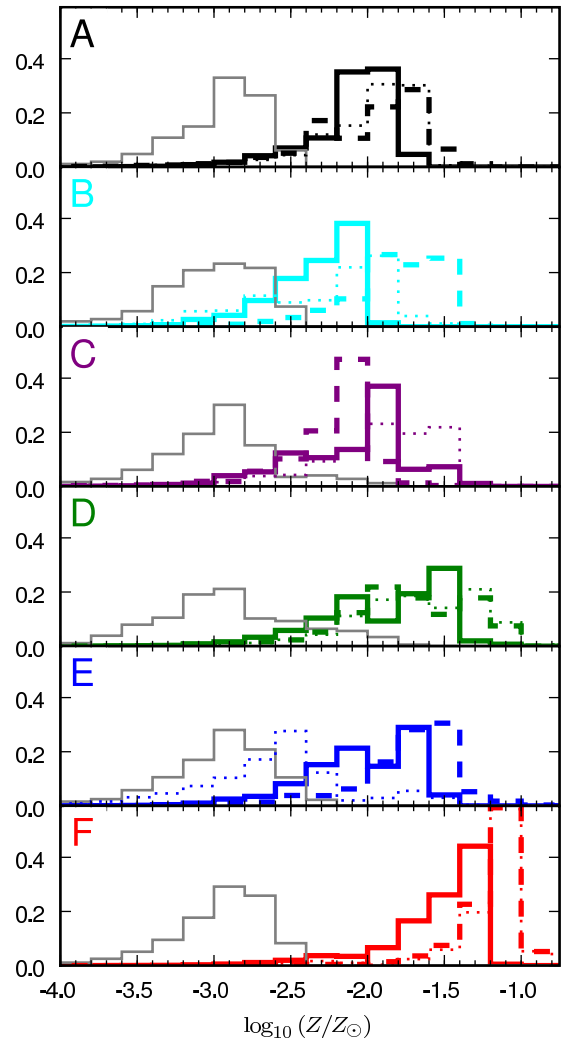}
\caption{Metallicity distribution functions of bright ($M_{\rm{V}} <
  -10$; solid coloured) and faint ($-10<M_{\rm{V}} < -5$; solid grey)
  satellites, halo stars in the `Solar shell' (dashed) and the entire
  halo ($3<r<280\;\rm{kpc}$, dotted). $Z$ is the total mass fraction
  of all metals.}
\label{fig:mdfs}
\end{figure}

\fig{fig:mdfs} shows luminosity-weighted metallicity distribution
functions (MDFs) for two selections of halo stars: a `Solar shell'
($5<r<12\,\rm{kpc}$; dashed lines) and the entire halo as defined in
\mnsec{sec:defining_haloes} (dotted). We compare these to MDFs for
stars in the surviving satellites in each halo, separating bright
($M_{\rm{V}} < -10$, $r < 280\,\rm{kpc};$ thick, coloured) and
`faint' ($-10<M_{\rm{V}}<-5$; thin, grey) subsets. All distributions
are normalized individually to the total luminosity in their sample
of stars.

The MDF of Solar-shell halo stars is typically broad, and tends to
peak at slightly higher metallicity (by $<0.5$ dex) than the
aggregated surviving bright satellites. The halo as a whole is
comparable to the Solar shell. A clear disparity is only evident in
Aq-E, where the halo appears to reflect more closely the distribution
of fainter, lower-metallicity satellites. In all cases, the MDF of
these faint satellites peaks at considerably lower metallicity than in
the halo or brighter satellites. \movt{We find that the `average'
  halo has an equivalent number of very metal-poor stars to the
  surviving bright satellites, although there are clear exceptions in
  individual cases. The fainter satellites have a substantially
  greater fraction of very metal-poor stars, in accordance with their
  low mean metallicities. Surviving satellites contain a greater
fraction of moderately metal-poor stars
($\log_{10}(Z/\rm{Z_{\sun}})<-2.5$) than the halo.}


\movt{Our halo models suggest that similar numbers of
  comparably luminous (and hence metal-rich) satellites contribute to
  the bright end of both the halo-progenitor and the
  surviving-satellite luminosity functions, and that these bright
  satellites are the dominant contributors to the halo. This supports
  the view that halo MDFs should resemble those of bright survivor
  satellites in their metal-poor tails. At very low metallicities the
  halo is dominated by the contribution of low-luminosity satellites
  which are exclusively metal-poor; the stars associated with these
  faint contributors are expected to represent only a very small
  fraction of the total halo luminosity.}

\begin{figure}
\includegraphics[width=84mm,clip=True, trim=0mm 0cm 0cm 0cm]{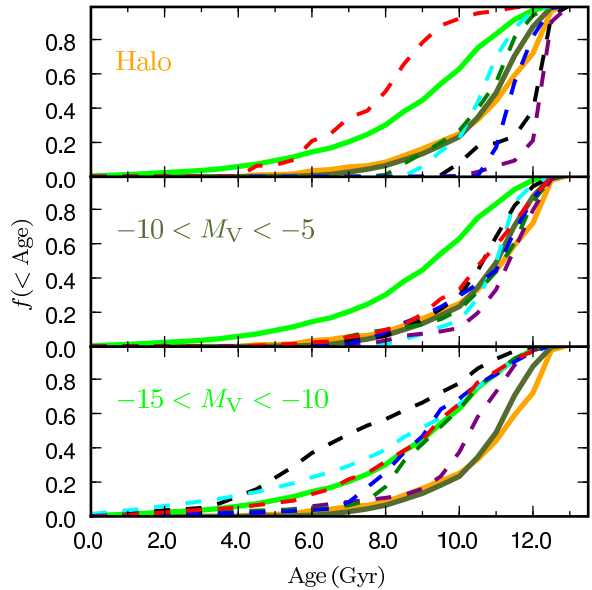}
\caption{The cumulative luminosity-weighted age distribution
  \newt{(mean of all six simulations)} for halo stars in the Solar
  shell (\newt{$5 < r < 12$ kpc, }orange, top \newt{panel}) compared
  to bright (\newt{$-15 <M_{\rm{V}} < -10$; } light green, bottom) and
  faint (\newt{$-10 < M_{\rm{V}} < -5$; }dark green, centre)
  satellites ($M_{\rm{V}} < -10$), showing individual contributions
  from each halo (dashed\newt{, colours as in previous figures})
  \newt{to the mean value represented by} each panel. The
total stellar masses of these three components over all haloes are
$1.04\times10^{9}$, $7.45\times10^{8}$ and
$3.45\times10^{8}\,\rm{M_{\sun}}$, respectively. }
\label{fig:agedistribution}
\end{figure}

Finally, \fig{fig:agedistribution} compares the luminosity-weighted
age distributions of halo stars in the Solar shell with those in the
surviving satellites ($M_{\rm{V}} < -5$), separated into bright and
faint subsets. \newt{The
  average of all six haloes} contains essentially no stars younger
than 5~Gyr (if we exclude halo Aq-F, which is strongly influenced by
the late accretion of an SMC-like object, this minimum age rises to
8~Gyr). The median age of halo stars is $\sim11$ Gyr. By contrast, the
brightest satellites have a median age of $\sim8$ Gyr and a
substantial tail to young ages (with $\sim20\%$ younger than 4 Gyr and
$\sim90\%$ younger than the median halo age). The distribution of old
stars in the faintest surviving satellites is similar to that of the
halo.

The true age distribution of halo stars is poorly constrained in
comparison to that of the satellites
(e.g. Tolstoy, Hill \& Tosi 2009). By comparing the colour and
metallicity distributions of Milky Way halo stars to those of the
Carina dSph, Unavane, Wyse \& Gilmore (1996) have argued that similar satellites
(i.e. those with a substantial fraction of intermediate-age stars)
could not contribute more than $\sim1\%$ to the halo (equivalent to a
maximum of $\sim60$ halo progenitors of Carina's luminosity). A
corresponding limit of $\leq6$ Fornax-like accretions in the last
$\sim10$ Gyr was derived from an analysis of higher metallicity stars
by the same authors, consistent with the progenitor populations of our
simulated stellar haloes.

It is important in this context that the satellites themselves form
hierarchically. In our models, between ten and twenty progenitors
are typical for a (surviving) galaxy of stellar mass comparable to
Sagittarius, or five to ten for a Fornax analogue. Satellites in
this mass range are the most significant contributors to our stellar
haloes. Their composite nature is likely to be reflected in their
stellar population mix and physical structure, which could
complicate attempts to understand the halo `building blocks' and the
surviving satellites in terms of simple relationships between mass,
age and metallicity.

\section{Conclusions}
\label{sec:conclusion}

We have presented a technique for extracting information on the spatial
and kinematic properties of galactic stellar haloes that combines a very
high resolution fully cosmological \lcdm{} simulation with a
semi-analytic model of galaxy formation. We have applied this technique
to six simulations of isolated dark matter haloes similar to or slightly
less massive than that of the Milky Way, adopting a fiducial set
of paramter values in the semi-analytic model \galform{}. The
structural properties of the surviving satellites have been used as a
constraint on the assignment of stellar populations to dark matter. We
found that this technique results in satellite populations and stellar
haloes in broad agreement with observations of the Milky Way and M31, if
allowance is made for differences in dark halo mass.

Our method of assigning stellar populations to dark matter particles
is, of course, a highly simplified approach to modelling star
formation and stellar dynamics. The nature of star formation in
dwarf galaxy haloes remains largely uncertain. In future,
observations of satellites interpreted alongside high-resolution
hydrodynamical simulations will test the validity of approaches such
as ours. As a further simplification, our models do not account for
a likely additional contribution to the halo from scattered {\em in
situ} (disc) stars, although we expect this contribution to be
minimal far from the bulge and the disc plane. The results outlined
here therefore address the history, structure and stellar
populations of the accreted halo component in isolation.

Our results can be summarised as follows:

\begin{itemize}
\item Our six stellar haloes are predominantly built by satellite
  accretion events occurring between $1<z<3$. They span a
  range of assembly histories, from
  `smooth' growth (with a number of roughly equally massive
  progenitors accreted steadily over a Hubble time) to growth in one
  or two discrete events.
\item Stellar haloes in our model are typically built from fewer than
  5 significant contributors. These significant objects have stellar
  masses comparable to the brightest classical dwarf spheroidals of
  the Milky Way; by contrast, fewer faint satellites contribute to the
  halo than are present in the surviving population.
\item Typically, the most massive halo contributor is accreted at a
  lookback time of between 7 and 11 Gyr ($z\sim1.5-3$) and deposits
  tidal debris over a wide radial range, dominating the contribution
  at large radii. Stars stripped from progenitors accreted at even
  earlier times usually dominate closer to the centre of the halo.
\item A significant fraction of the stellar halo consists of stars
  stripped from individual \textit{surviving} galaxies, contrary to
  expectations from previous studies (e.g. Bullock \& Johnston 2005). It
  is the most recent (and significant) contributors that are likely
  to be identifiable as surviving bound cores. Such objects have
  typically lost $\sim90\%$ of their original stellar mass.
\item We find approximately power-law density profiles for the
  stellar haloes in the range $10<r<100$ kpc. Those haloes formed by a
  superposition of several comparably massive progenitors have slopes
  similar to those suggested for the Milky Way and M31 haloes, while
  those dominated by a disproportionally massive progenitor have
  steeper slopes.
\item Our haloes have strongly prolate distributions of stellar mass
  in their inner regions ($c/a\sim0.3$), with one exception, where an
  oblate, disc-like structure dominates the inner $10-20$ kpc.
\item Haloes with several significant progenitors show little or no
  radial variation in their mean metallicity ($Z/\mathrm{Z_{\sun}}$) 
  up to 200~kpc. Those \newt{in which} a small number of progenitors
  dominate show stronger metallicity gradients over their full extent
  or sharp transitions between regions of different metallicity. The
  centres of these haloes are typically more enriched than their outer
  regions.
\item The stellar populations of the halo are likely to be chemically
  enriched to a level comparable to that of the bright surviving
  satellites, but to be as old as the more metal-poor surviving `ultra
  faint' galaxies. The very metal-poor tail of the halo distribution
  is dominated by contributions from a plethora of faint galaxies that
  are insignificant contributors to the halo overall.

\end{itemize}

\section*{Acknowledgments}

The simulations for the Aquarius Project were carried out at the
Leibniz Computing Centre, Garching, Germany, at the Computing Centre
of the Max-Planck-Society in Garching, at the Institute for
Computational Cosmology in Durham, and on the `STELLA' supercomputer
of the LOFAR experiment at the University of Groningen.

APC is supported by an STFC postgraduate studentship, acknowledges
support from the Royal Astronomical Society and Institute of Physics,
and thanks the KITP, Santa Barbara, for hospitality during the early
stages of this work. He also thanks \newt{Annette Ferguson for helpful
  comments and} Ben Lowing for code to calculate ellipsoidal fits to
particle distributions. CSF acknowledges a Royal Society Wolfson
Research Merit Award. AH acknowledges support from a VIDI grant by
Netherlands Organisation for Scientific Research (NWO). AJB
acknowledges the support of the Gordon \& Betty Moore foundation. JW
acknowledges a Royal Society Newton Fellowship. GDL acknowledges
financial support from the European Research Council under the
European Community's Seventh Framework Programme (FP7/2007-2013)/ERC
grant agreement n.~202781. \newt{We thank the referee for their
  suggestions, which improved the presentation and clarity of the
  paper.}

{}

\bsp
 
\label{lastpage}
\end{document}